\documentclass[aps,prd,10pt,nofootinbib,twocolumn,superscriptaddress,floatfix,notitlepage]{revtex4-1}
\pdfoutput=1

\usepackage{graphicx}
\usepackage{amsmath,amsfonts,amssymb}
\usepackage{booktabs}    
\usepackage{multirow}    
\usepackage[dvipsnames]{xcolor}
\usepackage[breaklinks,colorlinks,urlcolor=MidnightBlue,citecolor=WildStrawberry,linkcolor=Fuchsia]{hyperref}
\usepackage{verbatim}
\usepackage{enumitem}
\usepackage{aas_macros}
\usepackage{subcaption}

\newcommand{\td}{{\rm d}}

\newcommand{\be}{\begin{equation}}
\newcommand{\ee}{\end{equation}}
\newcommand{\bea}{\begin{equation} \begin{aligned}}
\newcommand{\eea}{\end{aligned} \end{equation}}

\def\lsim{\mathrel{\raise.3ex\hbox{$<$\kern-.75em\lower1ex\hbox{$\sim$}}}}
\def\gsim{\mathrel{\raise.3ex\hbox{$>$\kern-.75em\lower1ex\hbox{$\sim$}}}}

\usepackage{array}
\newcolumntype{R}[1]{>{\raggedleft\let\newline\\\arraybackslash\hspace{0pt}}m{#1}}
\newcolumntype{L}[1]{>{\raggedright\let\newline\\\arraybackslash\hspace{0pt}}m{#1}}
\usepackage[vmargin=2cm,hmargin=2cm]{geometry}

\usepackage[capitalise]{cleveref}
\usepackage[activate={true,nocompatibility},final,kerning=true,factor=1100,stretch=10,shrink=10]{microtype}
\usepackage{academicons}
\usepackage{fontawesome5}
\definecolor{orcidlogocol}{rgb}{0.65, 0.807, 0.223}
\newcommand{\orcid}[1]{$\,$\href{https://orcid.org/#1}{\textcolor{orcidlogocol}{\faOrcid}}}

\begin{document}

\title{CMB anisotropies from cosmic (super)strings in light of ACT DR6}

\author{Anastasios Avgoustidis\orcid{0000-0001-7247-5652}}\email{anastasios.avgoustidis@nottingham.ac.uk}

\author
{Edmund J. Copeland\orcid{0000-0003-3959-6051}}\email{edmund.copeland@nottingham.ac.uk}

\author{Adam Moss\orcid{0000-0002-7245-7670}}\email{adam.moss@nottingham.ac.uk}

\author{Juhan Raidal\orcid{0009-0002-2600-3632}}
\email{juhan.raidal@nottingham.ac.uk}
\affiliation{School of Physics and Astronomy, The University of Nottingham, Nottingham, NG7 2RD, UK}

\begin{abstract}

We present updated constraints on cosmic string and superstring parameters derived from Cosmic Microwave Background (CMB) anisotropies. The constraints are obtained via Markov Chain Monte Carlo (MCMC) analyses of the full \textit{Planck} temperature and polarization data combined with the Atacama Cosmology Telescope (ACT) Data Release 6 (DR6). For ordinary cosmic strings, we constrain the string tension $G\mu$, the string wiggliness parameter $\alpha$, and the self-chopping efficiency $\tilde{c}$. For cosmic superstrings, we constrain the fundamental string tension $G\mu_F$, the string coupling $g_s$, and a parameter $w$ describing the volume of the compact extra dimensions. In both cases, we find significantly tighter bounds on the string tension compared to previous analyses, obtaining $2\sigma$ upper limits of $G\mu < 3.66\times10^{-8}$ and $G\mu_F < 1.38\times10^{-8}$. We also discuss the significant prior-dependence of these results. The computational pipeline used in this work, including a modified version of \texttt{CAMB} capable of computing CMB anisotropies sourced by any active network described via unequal-time correlators, is released publicly as \texttt{CAMBactive} \cite{Raidal_CAMBactive_CAMB_extension_2026}.  

\end{abstract}

\maketitle

\section{Introduction}

Research on cosmic strings and cosmic superstrings has been revitalized by recent gravitational-wave (GW) observations, which have enabled direct searches for string-like topological defects using both interferometers \cite{LIGOScientific:2021nrg} and pulsar-timing arrays \citep{NANOGrav:2023hvm, Avgoustidis:2025svu, EPTA:2023xxk, Bian:2022tju}. These searches currently constrain the cosmic string tension -- and the corresponding fundamental tension for cosmic superstrings -- to the range $G\mu (G\mu_F)\lesssim10^{-10}-10^{-12}$. A crucial assumption underlying these bounds is that all cosmic (super)string loops lose their energy wholly through GW emission. However, as has long been pointed out, especially in the case of cosmic strings, a competing decay channel into particle radiation is also expected (see for example \cite{Bevis:2006mj,Hindmarsh:2018wkp}). For global strings, loop decay into Goldstone bosons (axions) is expected \cite{Baeza-Ballesteros:2023say}, while particle radiation from gauge strings has been studied primarily in Abelian-Higgs models \cite{Kume:2024adn, Hindmarsh:2022awe,Baeza-Ballesteros:2024sny}. Because the fraction of loop energy lost to particle emission remains uncertain, the existing GW bounds for cosmic strings assume that loops decay purely into GWs. Although particle emission mechanisms for cosmic superstrings are less well understood, and are expected to be suppressed relative to gravitational channels due to their fundamental gravitational nature, similar caveats apply.

A way of constraining cosmic strings and superstrings independently of GW observations is to study the CMB anisotropies sourced by the active population of strings as their network evolves. Since these anisotropies depend mainly on the mass of the long string network, we can obtain robust bounds on the underlying string parameters despite unknowns in the energy loss mechanisms of the network. Although there remain differences between results obtained by assuming maximal coupling to gravity, as in this work, and those derived from, for example, Abelian-Higgs strings \cite{Charnock:2016nzm,Lizarraga:2016onn}, these differences are at the level of a factor of two or three, rather than many orders of magnitude as is the case for GW-derived bounds.

\begin{figure*}
    \centering
    \includegraphics[width=\textwidth]{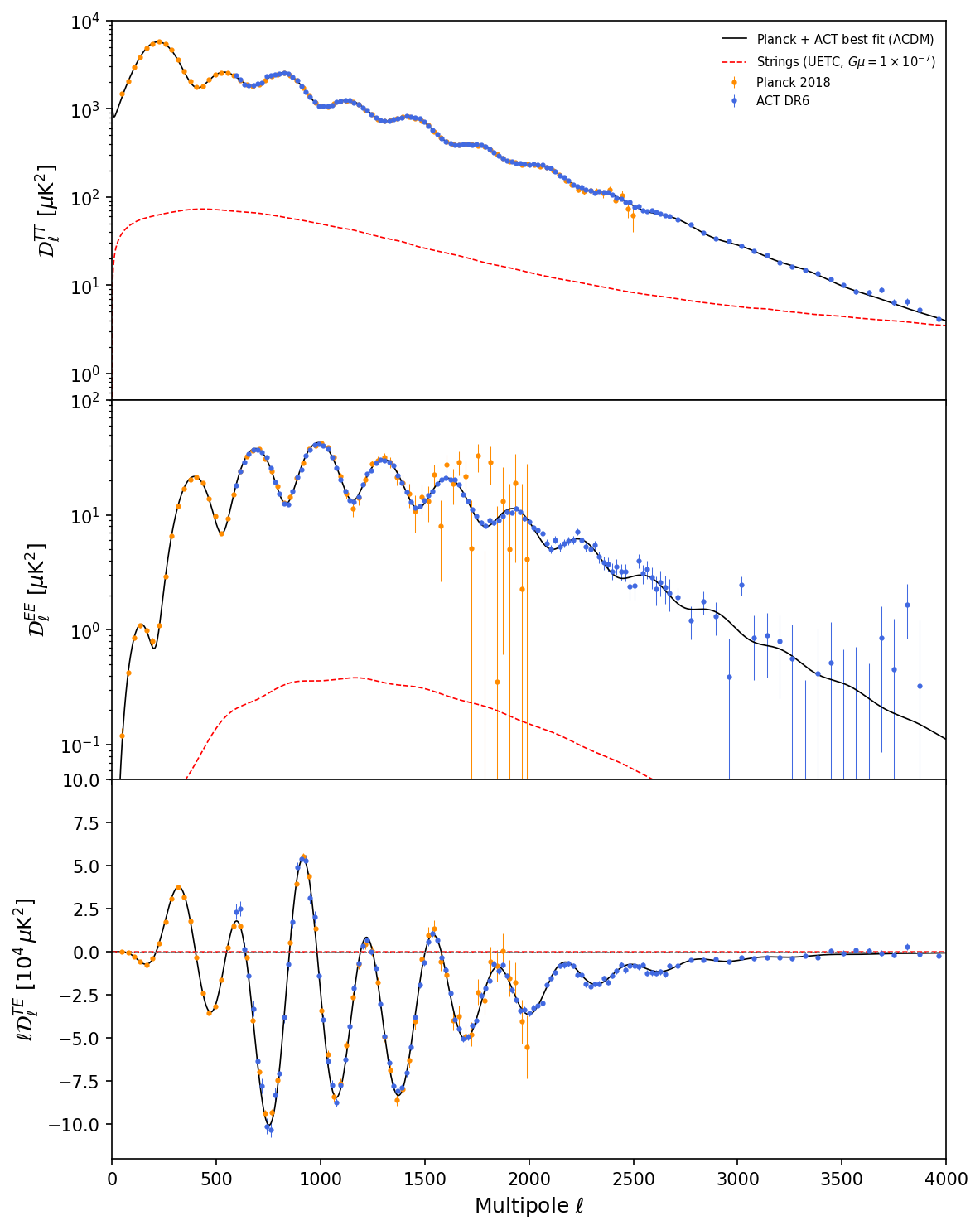}
    \caption{ACT DR6 and \textit{Planck} PR3 combined TT (top), EE (middle), and TE (bottom) power spectra. The solid black curve shows the ACT and \textit{Planck} best-fit $\Lambda$CDM model, while the dashed red curve shows cosmic strings from the USM model (Sec. \ref{sec:USM}) with $G\mu = 10^{-7}$, $\alpha = 1.9$, and $\tilde{c} = 0.23$.}
    \label{fig:power_spectrum}
\end{figure*}

Over the years cosmic (super)string networks have been increasingly constrained by CMB data, with current $2\sigma$ bounds for $\Lambda\mathrm{CDM}$+Nambu-Goto (super)strings sitting at $G\mu (G\mu_F)<1.1\times10^{-7}(2.8\times10^{-8})$. However, these constraints have not been updated since \citet{Charnock:2016nzm} and, with the full \textit{Planck} data \cite{Planck:2018vyg} and the high-resolution measurements from ACT DR6 \cite{AtacamaCosmologyTelescope:2025blo} now available, it is timely to revisit these bounds. At sufficiently high multipoles and string tension, contributions from cosmic (super)strings can dominate over $\Lambda$CDM. This is because the primary CMB anisotropies are exponentially suppressed at small scales by Silk damping, whereas the string-sourced signal, being continuously generated throughout cosmic history, does not suffer from this suppression and decays more slowly with $\ell$. The high-$\ell$ data from ACT is therefore expected to be particularly constraining, as illustrated in Fig.~\ref{fig:power_spectrum}.

In this work, we present an updated MCMC analysis of the CMB anisotropies sourced by cosmic strings and superstrings. The unequal-time correlators (UETCs) of the networks are computed using the unconnected segment model (USM) \citep{albrecht_evolution_1985,Pogosian:1999np}, supplemented by analytical approximations \citep{Avgoustidis:2012gb,Charnock:2016nzm}. These UETCs are then used as source terms to the linear Einstein-Boltzmann solver \texttt{CAMB} \citep{2011ascl.soft02026L} to compute the resulting CMB anisotropy spectra. Instead of the standard approach of precomputing CMB spectra on a grid of string parameters and interpolating between these values, we introduce neural-network (NN) emulators. We generate training datasets for both cosmic strings and superstrings, train NN models to emulate the anisotropy spectra, and integrate these emulators into \texttt{Cobaya} \citep{Torrado_2021} to obtain joint $\Lambda\mathrm{CDM}$+(super)string constraints via MCMC sampling. Importantly, any active source with a UETC description can be used in this pipeline. Our modified version of \texttt{CAMB}, capable of evolving arbitrary UETC-based active sources, is publicly available as \texttt{CAMBactive} \cite{Raidal_CAMBactive_CAMB_extension_2026}.

The structure of this paper is as follows. In Section~\ref{sec:modelling} we review the USM framework for computing the two-point correlators of cosmic (super)strings and summarize the analytical approximations used. Section~\ref{sec:calculation} describes the computational pipeline, including our novel implementation of NN emulators. In Section~\ref{sec:results} we present updated constraints on (super)string network parameters and discuss their implications. We conclude in Section~\ref{sec:conclusions}.

\section{Modelling of Correlators}
\label{sec:modelling}

Unlike the primordial perturbations generated during inflation, which serve purely as initial conditions, cosmic (super)strings are active sources that continuously generate perturbations throughout cosmic history. This continuous sourcing is incorporated by adding the Fourier components of the string network’s energy-momentum tensor as source terms in the linearized Einstein-Boltzmann equations used to compute CMB anisotropies. More specifically, these source terms are given by the eigenmodes of the network’s energy-momentum tensor's two-point unequal-time correlators (UETCs), defined as
\begin{align}
    \mathcal{C}_{\mu \nu ,\alpha \beta}(\mathbf{k},\tau,\tau')
    \equiv 
    \big\langle 
        \Theta_{\mu\nu}(\mathbf{k},\tau)\,
        \Theta^{*}_{\alpha\beta}(\mathbf{k},\tau')
    \big\rangle,
\end{align}
where $\Theta_{\mu\nu}(\mathbf{k},\tau)$ is the Fourier-space energy-momentum tensor of the string network.

\subsection{String Network Dynamics}

The evolution of a cosmic string network is governed by the dynamics of a population of long strings and the loops that they continuously produce via self-intersection. The standard approach to analysing cosmic string network dynamics analytically is the velocity-dependent one-scale (VOS) model, which assumes that the network can be, on large scales, treated as a collection of effectively straight segments whose statistical properties are described by a correlation length, $L$, and a root-mean-square velocity, $v$ \cite{Martins:2000cs}. In this approximation, $L$ is equal to both the inter-string separation and to the average length of string segments, enabling statistical averaging of the network's properties, which gives simple equations that govern the network parameters. In the VOS description for Nambu-Goto cosmic strings, the correlation length and velocity satisfy
\begin{align}
            \frac{\td \xi}{\td \tau}=&\phantom{+}\frac{1}{\tau}\bigg(v^2\xi\tau\frac{a'}{a}-\xi+\frac{\tilde{c}v}{2}\bigg),\label{VOSxi_comoving}\\
            \frac{\td v}{\td \tau}=&\phantom{+}(1-v^2)\bigg(\frac{k(v)}{\xi\tau}-2v\frac{a'}{a}\bigg)\,, \label{VOSv_comoving}
\end{align}
where $\xi$ is the correlation length in comoving coordinates (given by $\xi\tau=L/a$), $a$ is the scale factor, $\tau$ is conformal time and $a' \equiv da/d\tau$. To account for physics not captured by the large-scale averaging of the VOS approach, two parameters have been introduced -- the loop chopping efficiency, $\tilde{c}$, and the momentum parameter, $k(v)$. The loop chopping efficiency is a constant parameter that mediates the amount of energy lost to loops as the network evolves and strings intersect with each other. The momentum parameter encodes the effect of small-scale structure on the dynamics of string segments at the correlation-length scale, and has been phenomenologically found to have the form
\bea
    k(v)=\frac{2\sqrt{2}}{\pi}(1-v^2)(1+2\sqrt{2}v^3)\bigg(\frac{1-8v^6}{1+8v^6}\bigg).
\eea

The VOS model can be extended to cosmic superstring networks by considering multiple string types and the interactions between them. At formation, a superstring network contains light fundamental strings (F-strings) with tension $\mu_{\rm F}$, as well as heavier non-perturbative D-strings (one-dimensional Dirichlet branes) with tension $\mu_{\rm D}=\mu_{\rm F}/g_s$, where $g_s<1$ is the fundamental string coupling. These superstring ``species" can then combine to form bound states consisting of $p$ F-strings and $q$ D-strings, known as $(p,q)$-strings, which carry a conserved charge $(p,q)$ and have tension
\bea \label{pg-string-tension}
    \mu_{(p,q)}=\frac{\mu_{\rm F}}{g_s}\sqrt{p^2 g_s^2+q^2}\, .
\eea
It is convenient to assign a single index $i=1,2,\ldots$ to each superstring species. In this convention, the fundamental $(1,0)$ F-string is labelled $i=1$, the D-string $(0,1)$ as $i=2$, and the lightest bound state $(1,1)$ (the FD-string) as $i=3$, with higher $(p,q)$ states following accordingly. Although infinitely heavy bound states exist in principle, previous works have shown that restricting to $i=1,\ldots,7$ is sufficient to accurately capture the total evolution of the superstring network \citep{Avgoustidis:2009ke,Avgoustidis:2025svu,Avgoustidis:2012gb}.

The VOS equations describing cosmic superstrings are
\begin{widetext}

            \begin{align}
            \frac{\td \xi_i}{\td \tau}=&\phantom{+}\frac{1}{2\tau}\bigg[2v_i^2\xi_i\tau\frac{a'}{a}-2\xi_i+c_iv_i+\sum_{a,b}\bigg(\frac{d_{ia}^b\bar{v}_{ia}\xi_i\ell^b_{ia}}{\xi_a^2}-\frac{d^i_{ab}\bar{v}_{ab}\xi_i^3\ell^i_{ab}}{2\xi_a^2\xi_b^2}\bigg)\bigg],\label{eq:sxi}\\
            \frac{\td v_i}{\td \tau}=&\phantom{+}\frac{v_i^2-1}{\tau}\bigg[2v_i\tau\frac{a'}{a}-\frac{k_i}{\xi_i}-\sum_{a,b}b_{ab}^i\frac{\bar{v}_{ab}}{2v_i}\frac{(\mu_a+\mu_b-\mu_i)}{\mu_i}\frac{\xi_i^2\ell^i_{ab}}{\xi_a^2\xi_b^2}\bigg],\label{eq:svel}
            \end{align}
\end{widetext}
where $c_i$ is the {\it self-intersection} (or loop chopping) efficiency parameter for strings of type $i$, and $d^k_{ij}=d^k_{ji}$ are the {\it cross-string intersection} efficiency parameters describing processes in which a string of type $i$ interacts with a string of type $j$, with relative velocity $\bar{v}_{ij}$, to produce a type $k$ string segment of length $\ell_{ij}$. The coefficients, $b_{ab}^i$, determine whether the energy released during superstring interactions is transferred to the newly-formed superstring segment ($b_{ab}^i = d^i_{ab}$) or radiated away ($b_{ab}^i = 0$).

To understand the physical meaning of these terms, it is useful to view superstring interactions as ``zipping" processes. When two superstrings collide, they can join along a portion of their length to form a bound segment (a zipper), resulting in a trilinear vertex (Y-junction). Because the tension of the bound state is lower than the sum of the tensions of the incoming strings, energy is released during the process. The probability for two species $i$ and $j$ to form a junction depends on the kinematics of the collision. Cosmic superstrings typically have only mildly relativistic velocities, so the average relative velocity between colliding segments can be approximated as $\bar{v}_{ij}=\sqrt{v_i^2+v_j^2}$. The typical length of the zipper produced in the interaction is smaller than that of either parent segment and, following earlier works, we adopt the ansatz $\ell_{ij}^{-1}=L_i^{-1}+L_j^{-1}$,
where $L_i$ and $L_j$ are the characteristic correlation lengths of species $i$ and $j$, respectively.

Both the self- and cross- interaction coefficients, $c_i$ and $d^k_{ij}$, depend on a microphysical intercommuting probability ${\mathcal P}_{ij}$, which can be computed (or approximated) via various methods in string theory \cite{Jackson:2004zg,Hanany:2005bc}. Following \cite{Pourtsidou:2010gu}, we take 
\bea\label{eq:effinterc}
    c_i=\tilde{c}\times\mathcal{P}_i^{1/3}\ \ ,\ \ \ d_{ij}= \mathcal{P}_{ij}^{1/3}\, ,
\eea
where $\mathcal{P}_i \equiv \mathcal{P}_{ii}$ is the microphysical self-interaction probability for strings of type $i$, and $\tilde{c}$ is the self-interaction coefficient, chosen to equal the loop-chopping efficiency for cosmic strings. This is done such that, in the limit of maximal self-interaction probability, ${\mathcal P}_{i}=1$, we recover the standard cosmic string case with $c_i=\tilde{c}$. The matrix $d_{ij}$ is related to $d_{ij}^k$ by a kinematic factor $S^k_{ij}$ \cite{Avgoustidis:2009ke,Pourtsidou:2010gu} such that $d^k_{ij}=d_{ij}S^k_{ij}$. The kinematic matrix describes the probability that a given superstring collision produces a heavier -- $(p,q)+(p',q')\to (p + p', q + q')$ -- or lighter -- $(p,q)+(p',q')\to (p - p', q - q')$ -- string segment.

Finally, we must discuss the microphysical intercommutation probability, $\mathcal{P}_{ij}$. This quantity captures both the intrinsic quantum probability that two cosmic superstrings interact \cite{Jackson:2004zg,Hanany:2005bc}, and the probability that the strings ``miss" each other due to their motion in the compact extra dimensions \cite{Jones:2003da,Dvali:2003zj,Jackson:2004zg}. Following \cite{Pourtsidou:2010gu}, we write
\bea
    \mathcal{P}_{ij}=\mathcal{F}_{ij}(\bar{v}_{ij},\theta,g_s)\mathcal{V}_{ij}(w,g_s),
\eea
where the quantum interaction factor, $\mathcal{F}_{ij}$, depends on the relative velocity $\bar{v}_{ij}$ of the colliding strings, their collision angle $\theta$, and the fundamental string coupling $g_s$. The second factor, $\mathcal{V}_{ij}$, encodes the extra-dimensional effects on string dynamics and is a function of $g_s$ and a volume parameter $w$ \cite{Pourtsidou:2010gu}.

The parameter $w$ is defined as the ratio between the minimum volume $V_{\rm min}$ that a string segment can occupy in the extra dimensions (roughly the string thickness raised to the number of compact dimensions \cite{Dvali:2003zj}) and the total volume $V_{\rm FF}$ that the fundamental strings explore as they oscillate in those dimensions. The latter is bounded above by the full volume of the compact dimension \cite{Jackson:2004zg}, making $w$ an indirect lower bound on the size of the compact space. When $w=1$ the impact of extra dimensions is negligible, whereas small $w$ corresponds to the strings sampling a larger effective volume, which reduces the intercommutation probability. Further discussion of the physical meaning of $w$ can be found in \cite{Pourtsidou:2010gu,Jackson:2004zg,Jackson:2007hn}. A more detailed treatment of $\mathcal{P}_{ij}$, including explicit forms for $\mathcal{F}_{ij}$ and $\mathcal{V}_{ij}$, is provided in \cite{Pourtsidou:2010gu}.

\subsection{Unconnected Segment Model}
\label{sec:USM}

Accurately computing the energy-momentum tensor of a (super)string network, in principle, requires a full network simulation, tracking the positions and evolution of individual strings over long time-scales. This approach, however, is computationally unfeasible, and analytic approaches or simplifications are required. Following the logic of the VOS model, the string network can be approximated as a collection of unconnected string segments of length $L$ moving with speed $v$, both of which are determined by the VOS equations. In this framework, known as the unconnected segment model (USM) originally introduced by \citet{Albrecht:1997mz} and \citet{Pogosian:1999np}, analytical expressions can be found for both the energy-momentum tensor and its correlators.

In the USM, the network is first divided into correlation-length segments. A further simplification is then made by consolidating all string segments that decay between two given times, so that only populations decaying at distinct times need be tracked. The number of strings decaying between times $\tau_{i-1}$ and $\tau_i$ in a volume $V$ is
\bea
    N_{\rm d}(\tau_i) = V\,[n(\tau_{i-1}) - n(\tau_i)],
\eea
where $n(\tau) = C(\tau)/(\xi \tau)^3$ is the string number density at time $\tau$. The time-dependent factor $C(\tau)$ is chosen such that $n(\tau) \propto 1/(\xi \tau)^3$ at all times. As discussed by \citet{Charnock:2016nzm}, in the analytic limit where the number of string segments becomes infinite, $C(\tau) \to 1$.

Knowing the number of string segments in each consolidated string segment enables us to write down the energy-momentum tensor of the string network as the sum
\bea\label{eq:USMtensor}
    \Theta_{\mu\nu}=\sum^K_{i=1}\sqrt{N_{\rm d}(\tau_i)}\Theta_{\mu\nu}^i T^{\rm off}(\tau,\tau_i,L_f),
\eea
where $K$ is the number of consolidated string segments and $T^{\rm off}(\tau,\tau_i,L_f)$ is a decay factor that removes the contribution of the $i^{\rm th}$ consolidated string segment after it has decayed. The parameter $0<L_{\rm f}\leq1$ determines the steepness of the turn off. Following \cite{Charnock:2016nzm}, the decay factor takes the functional form
\bea\label{eq:Toff}
    T^{\rm off}(\tau,\tau_i,L_{\rm f})=\left\{\begin{array}{cl}1&\tau<L_{\rm f}\tau_i\\1/2+1/4(y^3-3y)&L_{\rm f}\tau_i<\tau<\tau_i\\0&\tau_i<\tau\end{array}\right.,
\eea
where
\bea
    y=\frac{2\ln(L_{\rm f}\tau_i/\tau)}{\ln(L_{\rm f})}-1\,.
\eea
Note that the time steps in the USM are determined by the number of consolidated string segments. As this tends to infinity, the time between consolidated string decays becomes infinitesimal and $L_{\rm f}\to1$, making $T^{\rm off}$ approach a Heaviside theta function.

\subsection{Correlators}

Using Eq.~\eqref{eq:USMtensor}, the unequal-time correlator can be computed analytically by integrating over all string network configurations: 
\begin{widetext}
    \bea
        \langle \Theta(k,\tau_1) \Theta(k,\tau_2) \rangle
        = \frac{2}{16\pi^3} f(\tau_1,\tau_2,\xi,L_{\rm f}) 
        \int_0^{2\pi} \mathrm{d}\phi \int_0^\pi \sin{\theta} \, \mathrm{d}\theta 
        \int_0^{2\pi} \mathrm{d}\psi \int_0^{2\pi} \mathrm{d}\chi \, 
        \Theta(k,\tau_1) \Theta(k,\tau_2),
    \eea
\end{widetext}
where $\Theta$ now represents a component of the energy-momentum tensor: the density mode $\Theta_{00}$, the anisotropic stress mode $\Theta_S$, the vector mode $\Theta_V$, or the tensor mode $\Theta_T$. The angles $(\phi,\theta,\psi)$ and the phase $\chi$ specify the spatial and velocity orientations of the string segment within the chosen coordinate system (see \cite{Avgoustidis:2012gb,Charnock:2016nzm} for details). The scaling factor is given, in the limit of $L_{\rm f}\to1$, by
\bea
    f(\tau_1,\tau_2,\xi,L_{\rm f})&=\sum_{i=1}^K N_{\rm d}(\tau_i)T^{\rm off}(\tau_1,\tau_i,L_{\rm f})T^{\rm off}(\tau_2,\tau_i,L_{\rm f})\\
    &=\frac{1}{(\xi(\max{[\tau_1,\tau_2]})\max{[\tau_1,\tau_2]})^3}.
\eea
Note that the above expressions are valid for a single string type. For cosmic superstrings, each string species satisfies these equations separately.

It is possible to analytically perform the integrals over $\phi$, $\psi$ and $\chi$, leaving only the integral over $\theta$, which can, in turn, be expressed as a sum of Bessel functions \cite{Avgoustidis:2012gb}. This gives the simplified, but still general expression for the energy-momentum correlator in the USM as
\bea\label{eq:UETCanalytic}
    \langle\Theta(k,\tau_1)\Theta(k,\tau_2)\rangle=&\frac{f(\tau_1,\tau_2,\xi,L_{\rm f})\mu^2}{k^2(1-v^2)}\times\\
    &\sum_{i=1}^6 A_i[I_i(x_-,\rho)-I_i(x_+,\rho)],
\eea
where $\rho=k|\tau_1-\tau_2|v$, $x_{1,2}=k\xi\tau_{1,2}$ and $x_{\pm}=(x_1\pm x_2)/2$. Both the integral identities, $I_i$, and the amplitude $A_i$ for each scalar, vector and tensor UETC component can be found in Tables I and II of \cite{Charnock:2016nzm}. 

In this work we are also allowing for the possibility of wiggly strings, in which, unlike for Nambu-Goto strings, the mass per unit length, $U$ and tension, $T$ of the strings no longer follow $U=T=\mu$. To model this, we follow \cite{Carter:1990nb}, (see also \cite{Pogosian:1999np}) and introduce the string wiggliness parameter, $\alpha$, which defines the effective mass per unit length, $\tilde{U}$ and tension, $\tilde{T}$ of a wiggly string through 
\be
    \tilde{U}=\alpha\mu, \tilde{T}=\mu/\alpha.
\ee
The integral identities in Eq.~\eqref{eq:UETCanalytic} depend heavily on $\alpha$, and it is one of the model parameters that will be constrained in Sec. \ref{sec:results}.

The full analytic expression in Eq.~\eqref{eq:UETCanalytic} can be simplified significantly in certain regimes. In the equal time correlator (ETC) case, $\tau_1=\tau_2=\tau$, $x_1=x_2=x_+=0$ and $x_-=\rho=0$, the exact analytic expression becomes
\bea\label{eq:ETCanalytic}
    \langle\Theta(k,\tau)\Theta(k,\tau)\rangle=\frac{f(\tau,\tau,\xi,L_{\rm f})\mu^2}{k^2(1-v^2)}C,
\eea
while in the small $x$ limit, an approximation can be found of the form
\bea\label{eq:UETCsmallx}
    \langle\Theta(k,\tau_1)\Theta(k,\tau_2)\rangle=\frac{f(\tau_1,\tau_2,\xi,L_{\rm f})\mu^2}{k^2(1-v^2)}B,
\eea
with the coefficients $B$ and $C$ for each correlator mode also available in Table II of \cite{Charnock:2016nzm}.
The above expressions enable the fast analytic calculation of the UETCs in the USM and will be used throughout the computations that follow.

\section{CMB anisotropies}
\label{sec:calculation}
\subsection{\texttt{CAMBactive}}

Although cosmic strings are incoherent sources of CMB perturbations, individual eigenmodes of the UETCs are coherent. This allows them to be used as source terms in the Einstein-Boltzmann equations solved by \texttt{CAMB} \cite{2011ascl.soft02026L}, the industry-standard code for calculating CMB anisotropies.

Using the analytic formulae of Section~\ref{sec:modelling}, we compute the UETC for a given wavenumber $k$ on a discrete $n \times n$ grid of logarithmic values of $k\tau_1$ and $k\tau_2$. The resulting matrix is then diagonalised as
\bea\label{eq:decomp}
    &(k^2 \tau_1 \tau_2)^\gamma \sqrt{\tau_1 \tau_2} \langle \Theta(k,\tau_1) \Theta(k,\tau_2) \rangle \nonumber\\
    &\quad= \sum_{i=1}^n \lambda_i \, u_i(k\tau_1) \otimes u_i(k\tau_2),
\eea
where $\lambda_i$ and $u_i$ are the eigenvalues and eigenvectors, respectively. The scaling factor $(k^2 \tau_1 \tau_2)^\gamma$ allows for a more
efficient reconstruction of the UETC when the eigenmodes are truncated below $n$. The choice $\gamma = 0.25$
gives the best reconstruction on scales that give the
dominant contribution to the CMB anisotropies.

The two scalar modes of the string energy-momentum tensor, $\Theta_{00}$ and $\Theta_S$, are correlated and cannot be decomposed independently. In this case, the UETC is discretised on a $2n \times 2n$ grid, yielding a block matrix with density and stress self-correlations along the diagonal and the $00$-$S$ cross-correlations on the off-diagonal. Diagonalisation of this matrix produces $2n$ eigenmodes, with the first $n$ corresponding to the density and the remaining $n$ to the anisotropic stress.

After diagonalisation, the orthogonal eigenvectors are used in the \texttt{CAMB} linear Einstein-Boltzmann equations as source terms of the form $u_i(k\tau)/((k\tau)^\gamma\tau^{1/2})$, and the total power spectrum is obtained by summing the contributions from each eigenmode:
\bea
    C_\ell = \sum_{i=1}^{n} \lambda_i C_\ell^i.
\eea
Note that, for cosmic superstrings, this procedure is repeated separately for each superstring species, and the total power spectrum is obtained by summing the contributions from all relevant species. Since heavier bound states have smaller number densities, it is not necessary to calculate the power spectrum for all 7 superstring species required to get convergent VOS solutions, and only the three lightest species are included in the MCMC analysis \cite{Charnock:2016nzm}.

The method of using UETC eigenvectors as Einstein-Boltzmann sources is general and can be applied to any active source with calculable UETCs. As active sources are not implemented in the production version of \texttt{CAMB}, we have developed \texttt{CAMBactive}, a modified branch that allows the calculation of CMB anisotropies from arbitrary UETC eigenmodes. The codebase also provides examples for computing cosmic string UETCs and their corresponding CMB power spectra.

\subsection{Power spectrum resolution}

By virtue of the analytic forms of the UETCs outlined in Section~\ref{sec:modelling}, the correlator matrices can be computed at high resolution with reasonable computational cost. For example, on a 16-core Apple M3 GPU, evaluating the UETCs at a single wavenumber on a $1024\times1024$ grid takes only $\mathcal{O}(1)\,$s. As computing the CMB power spectrum using \texttt{CAMBactive} requires many evaluations of the UETCs at different wavenumbers, finding the full anisotropy spectrum is much slower. For example, for a $1024\times1024$ UETC grid the calculation can take several hours when all eigenmodes are included. As such, before performing a full MCMC analysis, we must determine the optimal UETC grid resolution and the minimum number of eigenmodes required to achieve an acceptably accurate approximation of the full power spectrum. Ideally, the resolution errors should be negligible with respect to cosmic variance (CV) noise at the string tensions we expect to constrain.

To establish a reference for comparison, we first construct a baseline high-resolution ``true'' spectrum. This is obtained by progressively increasing the UETC grid density and computing the resulting CMB power spectra with all eigenmodes included. As the resolution increases, differences between successive spectra eventually become negligible relative to cosmic variance noise, given by
\be
    \Delta C_\ell^2=\frac{2}{2\ell+1}\left(C_\ell^{\rm tot}\right)^2,
\ee
where $C_\ell^{\rm tot}$ is the total power spectrum including both the $\Lambda$CDM and string contributions. We quantify the convergence by computing the signal-to-noise ratio (SNR) of the difference between spectra. Using this procedure, we adopt as baselines a $1200\times1200$ grid for cosmic strings and a $1024\times1024$ grid for cosmic superstrings with all eigenmodes included in both cases. Any spectrum whose SNR relative to these baselines remains below detectability is deemed sufficiently accurate for use in the MCMC analysis.

\begin{figure*}
    \centering
    \includegraphics[width=\textwidth]{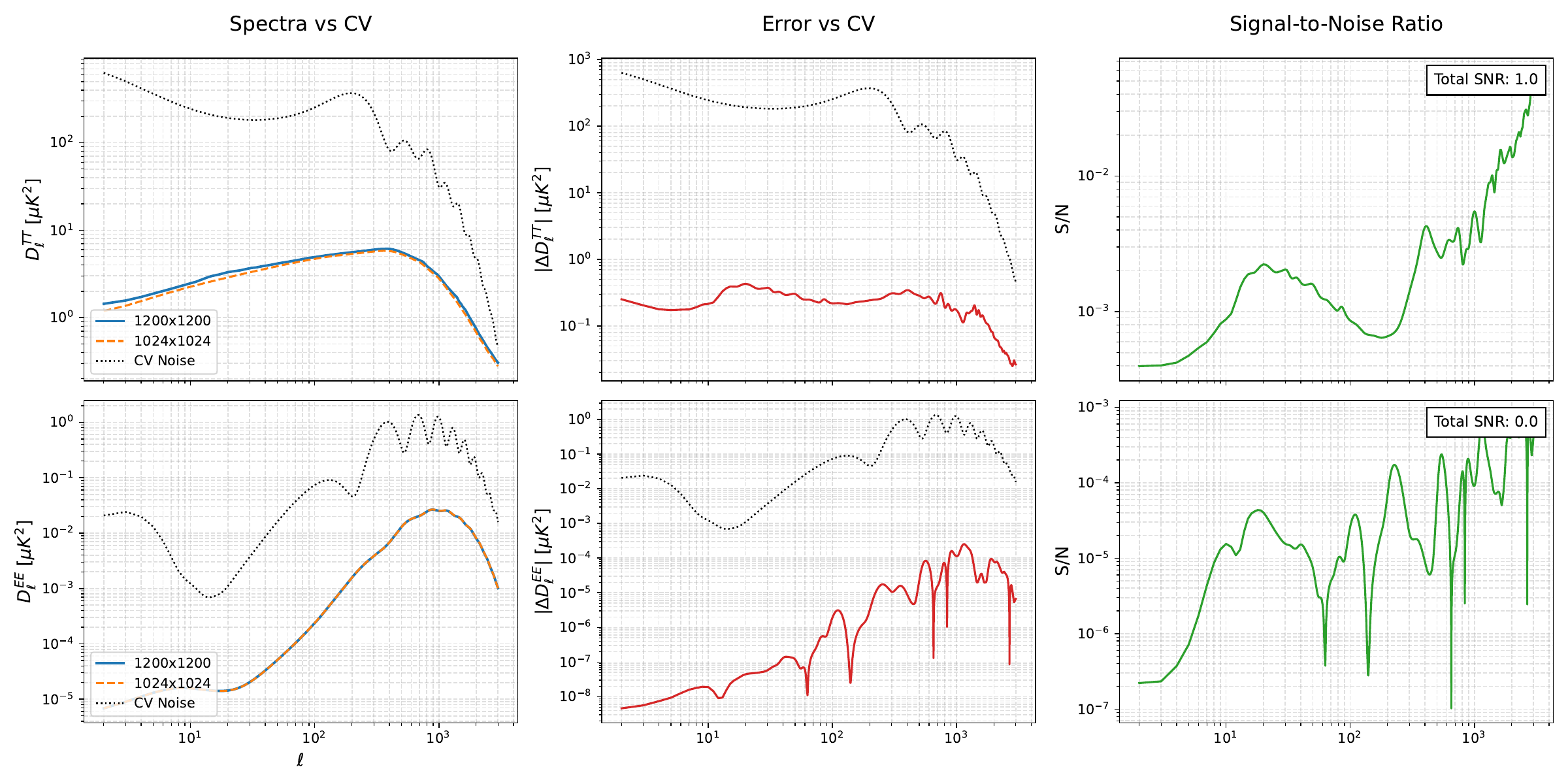}
    \caption{Cosmic string anisotropy spectra calculated from UETCs with different grid resolutions. The first column shows the high-resolution ``true'' spectrum, the chosen lower resolution spectrum and the CV noise. The second column shows the difference between the two spectra and the CV noise, while the third column gives the SNR at each multipole and the total SNR. The spectra are calculated using $G\mu=8\times 10^{-8}$. Only the TT and EE modes are shown; the remaining modes contribute subdominantly to the total SNR.}
    \label{fig:string_noise}
\end{figure*}

By gradually lowering the grid resolution and truncating the eigenmode sum, we find that, for cosmic strings, a $1024\times1024$ grid with $256$ eigenmodes yields, with respect to the baseline spectrum, a TT-mode CV SNR of $2.0$ at the previously determined $2\sigma$ upper bound on the string tension, $G\mu = 1.1\times10^{-7}$, and up to $\ell_{\rm max}=3000$. This is deemed sufficient, as reaching a SNR of 1, which happens at $G\mu = 8\times10^{-8}$, is within the expected improvement of the bounds. As we are using ACT DR6 high-multipole data, it is also important to check that our resolution is valid with respect to ACT noise. The SNR of the above resolution in the dominant TT-mode, with both ACT and CV noise, up to $\ell_{\rm max}=4500$, is 0.2, well below detectability. An example plot illustrating the above analysis can be seen in Fig.~\ref{fig:string_noise}.

Similar analysis for cosmic superstrings shows that a $1024\times1024$ grid with $100$ eigenmodes summed gives a TT-mode CV SNR at $\ell_{\rm max}=3000$ of 15.4 and a CV+ACT SNR at $\ell_{\rm max}=4500$ of 0.58 at the previously determined $2\sigma$ upper bound on the fundamental string tension, $G\mu_F = 2.8\times10^{-8}$. As superstrings have a stronger signature at high multipoles when compared to cosmic strings, we were unable to achieve resolution errors that are negligible w.r.t CV noise and computationally reasonable. Regardless, the above resolution is sufficient for the ACT-based MCMC analysis presented below.

\subsection{Power spectrum emulators}

For both cosmic strings and superstrings, evaluating the CMB power spectra with the resolutions determined above takes approximately five minutes per parameter set on a 16-core Apple M3 GPU. Although this is remarkably fast compared to network-simulation-based methods, it remains far too slow for direct use in an MCMC analysis, which typically requires tens of thousands of likelihood evaluations for convergence. Fortunately, the CMB power spectra sourced by cosmic (super)strings are smooth, well-behaved, and free of the multiple acoustic-peak structure characteristic of inflationary perturbations. This makes them ideal candidates for emulation with neural networks.

Conventional interpolation schemes could, in principle, also be used to approximate spectra between precomputed points in parameter space. However, reliable interpolation requires a dense grid of precomputed spectra, and the computational cost scales poorly with dimensionality -- each additional free parameter dramatically increases the number of precomputed spectra needed for good performance. Neural networks, by contrast, bypass many of these issues. We find that only $\mathcal{O}(1000)$ precomputed training spectra are sufficient to achieve median accuracies exceeding $95\%$ in both the cosmic string and cosmic superstring cases. This is particularly notable for superstrings, where the parameter space includes two additional dimensions, $(g_s, w)$, beyond the cosmic string case.

We implement the emulators as a fully connected neural network multilayer perceptron that maps the set of input parameters -- cosmological parameters + (super)string parameters -- directly to the full set of multipole spectra $\{C_\ell^{TT}, C_\ell^{EE}, C_\ell^{BB}, C_\ell^{TE}\}$. The network itself for both cosmic strings and superstrings consists of a shared backbone followed by two output ``heads'', one predicting the log-amplitudes of all spectra and one predicting the TE-mode signs. For both emulators we find that a relatively compact architecture of four hidden layers with widths [512, 512, 512, 256] and SiLU activations provides median accuracy (averaged over all multipoles) of $98.36\%$ and $95.04\%$ for the dominant TT-mode of cosmic strings and cosmic superstrings respectively.

\begin{figure*}
    \centering
    \includegraphics[width=\textwidth]{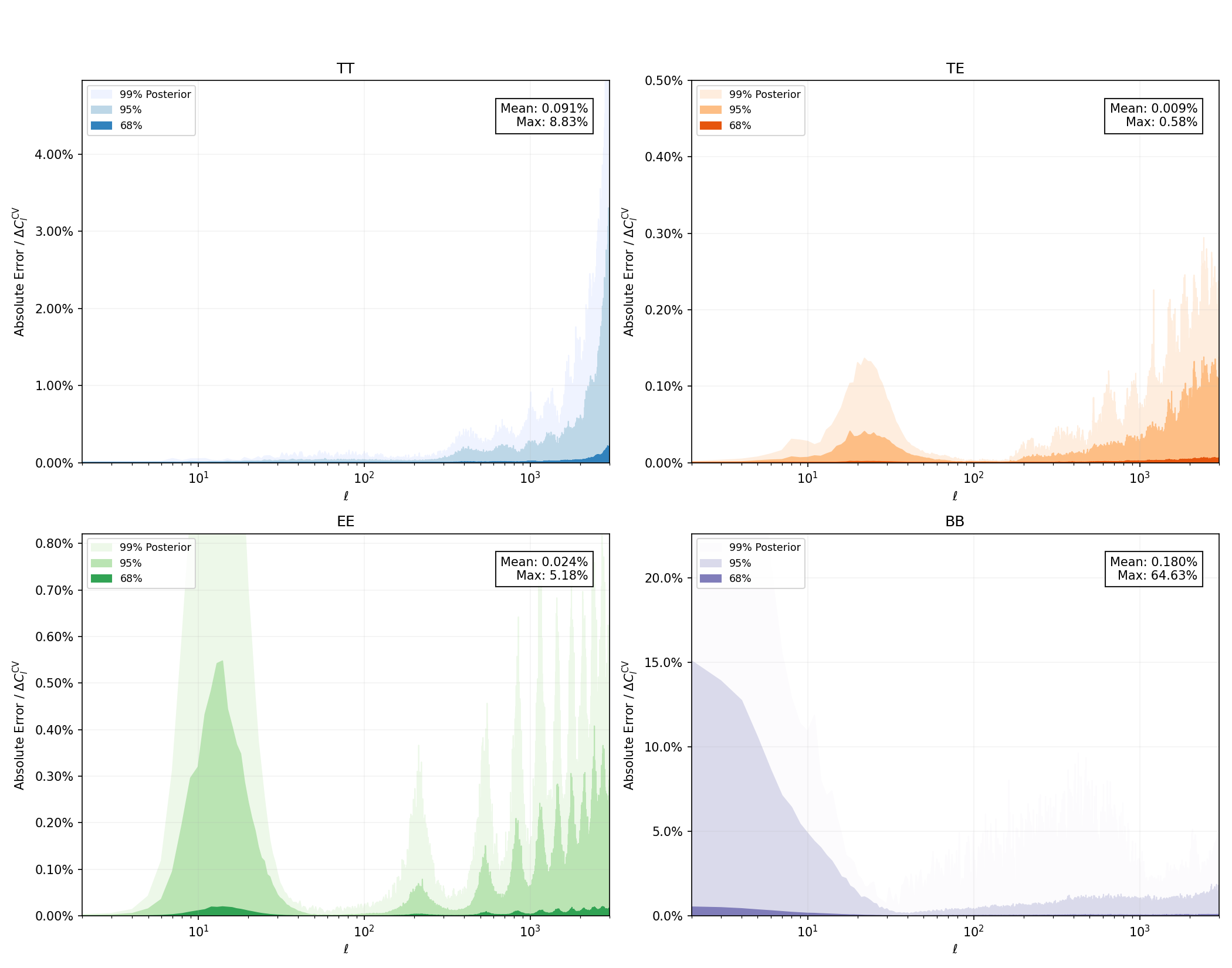}
    \caption{Fan plots showing the distribution of emulator errors from the true spectrum with respect to the cosmic variance error. The cosmological and string parameter values used are selected from MCMC chains and represent physical scenarios.}
    \label{fig:fan}
\end{figure*}

The raw accuracy values quoted above are a poor representation of the emulators' quality in practice, as they are obtained over a random sample of parameter values, which are not necessarily physical or compatible with data. To make sure that the emulators are accurate enough, we perform posterior validation by using the MCMC chains obtained via the simulations in Section~\ref{sec:results}. The parameter sets of these chains correspond to realistic scenarios favoured by the data, and if the emulators perform well here, then their use is justified. Using a random subset of 200 chains, we create a fan plot of the emulator errors relative to the $\Lambda$CDM baseline CV, which can be seen in Fig.~\ref{fig:fan} for cosmic strings. The dominant TT-mode sees a mean error of less than $0.1\%$, although there are larger errors present at higher multipoles, with the maximum error over the whole sample still being below $10\%$. Although not perfect, this performance is sufficient for our analysis, especially since cosmic (super)strings contribute subdominantly to the total CMB anisotropy spectrum.

For additional verification, we trained an alternative set of emulators on training data compressed via principal component analysis (PCA). Instead of predicting the full power spectra directly, these networks predict the coefficients of a truncated principal-component expansion, which were then used to reconstruct the spectra. This complementary approach yielded predictions that were consistent with those of the direct (non-PCA) emulator. The close agreement between the two strategies provides further evidence for the robustness and reliability of the emulator approach.

\section{Results and Discussion}
\label{sec:results}

The neural networks for the final MCMC analysis are trained on sets of $3000$ and $1000$ spectra for the cosmic strings and superstrings respectively. These emulators are integrated into a custom \texttt{Cobaya} \cite{Torrado_2021} theory class which, at each step of the MCMC, calculates the baseline $\Lambda$CDM anisotropy power spectrum using \texttt{CAMB}, and adds to it the (super)string spectra as predicted by the emulators. This combined spectrum is then used to obtain $\Lambda$CDM+strings constraints on both the standard cosmological parameters and the relevant stringy parameters, namely $\{G\mu,\alpha,\tilde{c}\}$ for cosmic strings and $\{G\mu_F,\alpha,\tilde{c},g_s,w\}$ for cosmic superstrings. For the latter, we assume that all string species share the same effective wiggliness.

For our \textit{Planck}-only analysis, we use the CamSpec PR4 likelihood \cite{Rosenberg:2022sdy} combined with the SRoll2 low-$\ell$ polarization likelihood \cite{Delouis:2019bub}. For the \textit{Planck}+ACT analysis, we use the ACT DR6 likelihood \cite{AtacamaCosmologyTelescope:2025blo}, which internally includes a cut version of the \textit{Planck} 2018 PR3 likelihood \cite{Planck:2019nip} at low multipoles, again combined with SRoll2 low-$\ell$ polarization.

\begin{figure*}[t]
    \centering
    \begin{subfigure}{0.48\textwidth}
        \includegraphics[width=\textwidth]{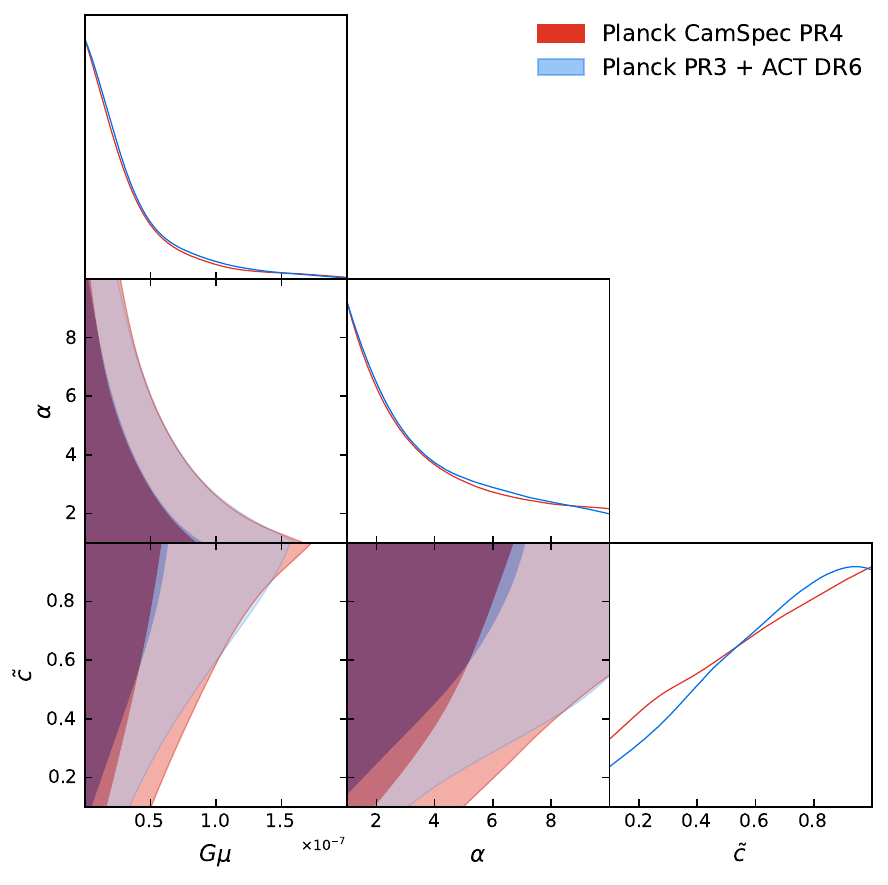}
        \caption{Linear $G\mu$, flat priors}
    \end{subfigure}
    \hfill
    \begin{subfigure}{0.48\textwidth}
        \includegraphics[width=\textwidth]{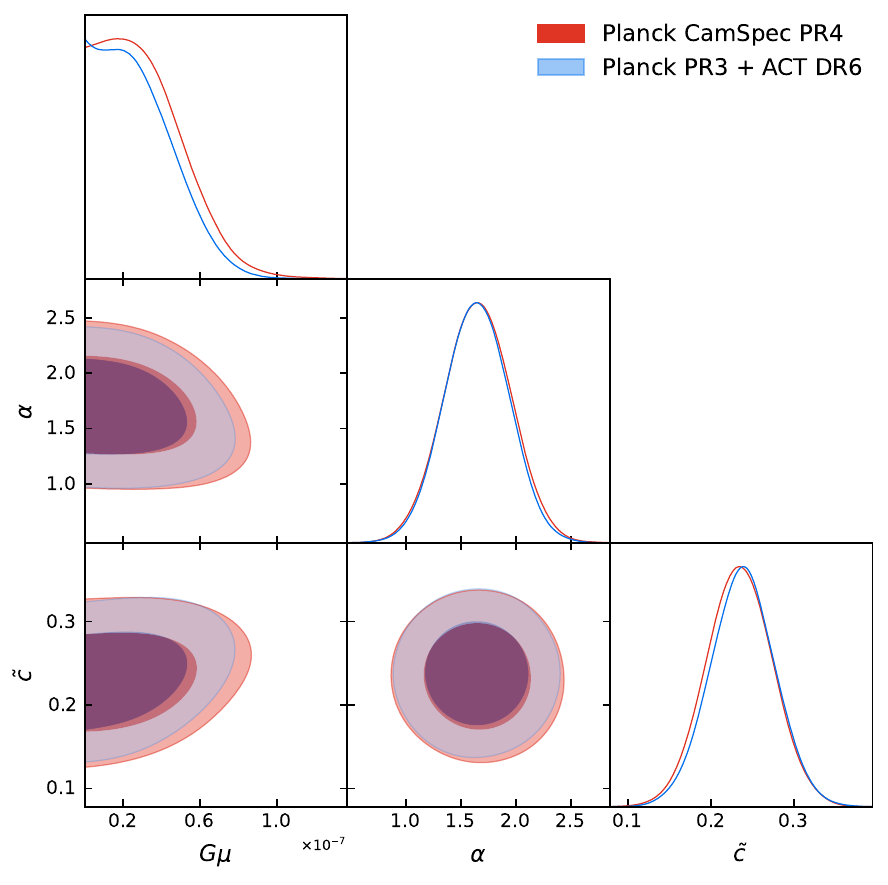}
        \caption{Linear $G\mu$, Gaussian priors}
    \end{subfigure}
    
    \vspace{0.5cm} 
    
    \begin{subfigure}{0.48\textwidth}
        \includegraphics[width=\textwidth]{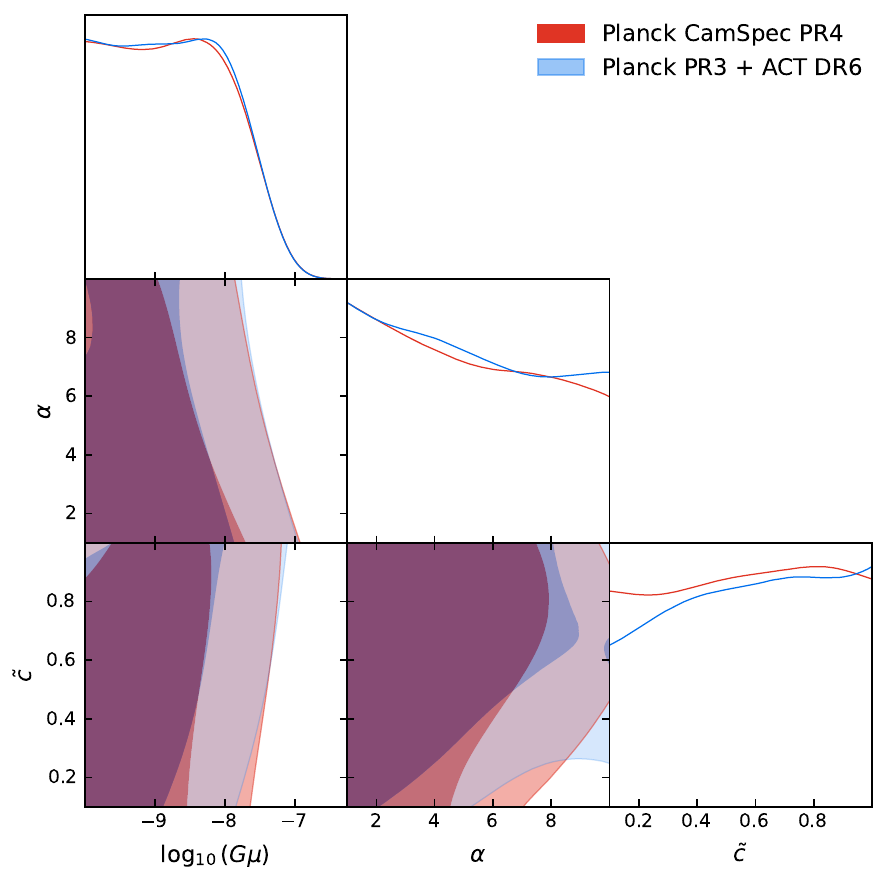}
        \caption{$\log_{10}G\mu$, flat priors}
    \end{subfigure}
    \hfill
    \begin{subfigure}{0.48\textwidth}
        \includegraphics[width=\textwidth]{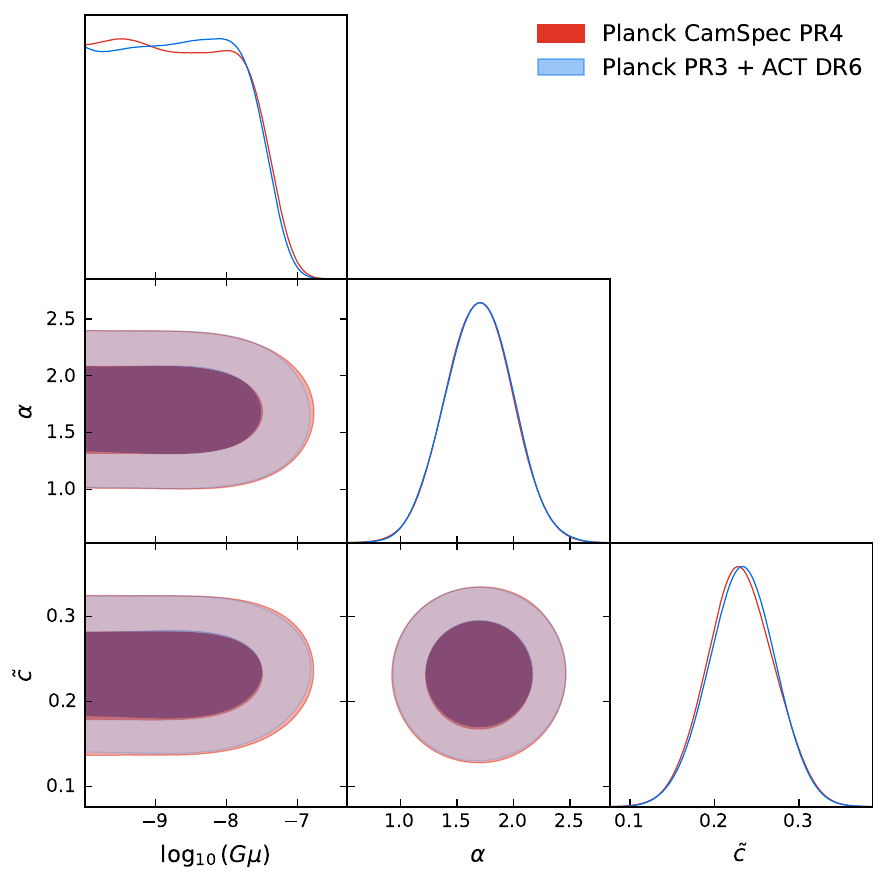}
        \caption{$\log_{10}G\mu$, Gaussian priors}
    \end{subfigure}
    
    \caption{Marginalized posterior distributions for the ordinary cosmic string parameters $\{G\mu, \alpha, \tilde{c}\}$. In each panel, we compare the constraints from \textit{Planck} CamSpec PR4 (red) and \textit{Planck} PR3 + ACT DR6 (blue). The left column shows the results under uninformative flat priors, highlighting the prior volume effect between linear (a) and log (c) parametrizations. The right column shows the results with physically-motivated Gaussian priors on $\alpha$ and $\tilde{c}$.}
    \label{fig:stringResults}
\end{figure*}

\begin{figure*}[t]
    \centering
    \begin{subfigure}{0.48\textwidth}
        \includegraphics[width=\textwidth]{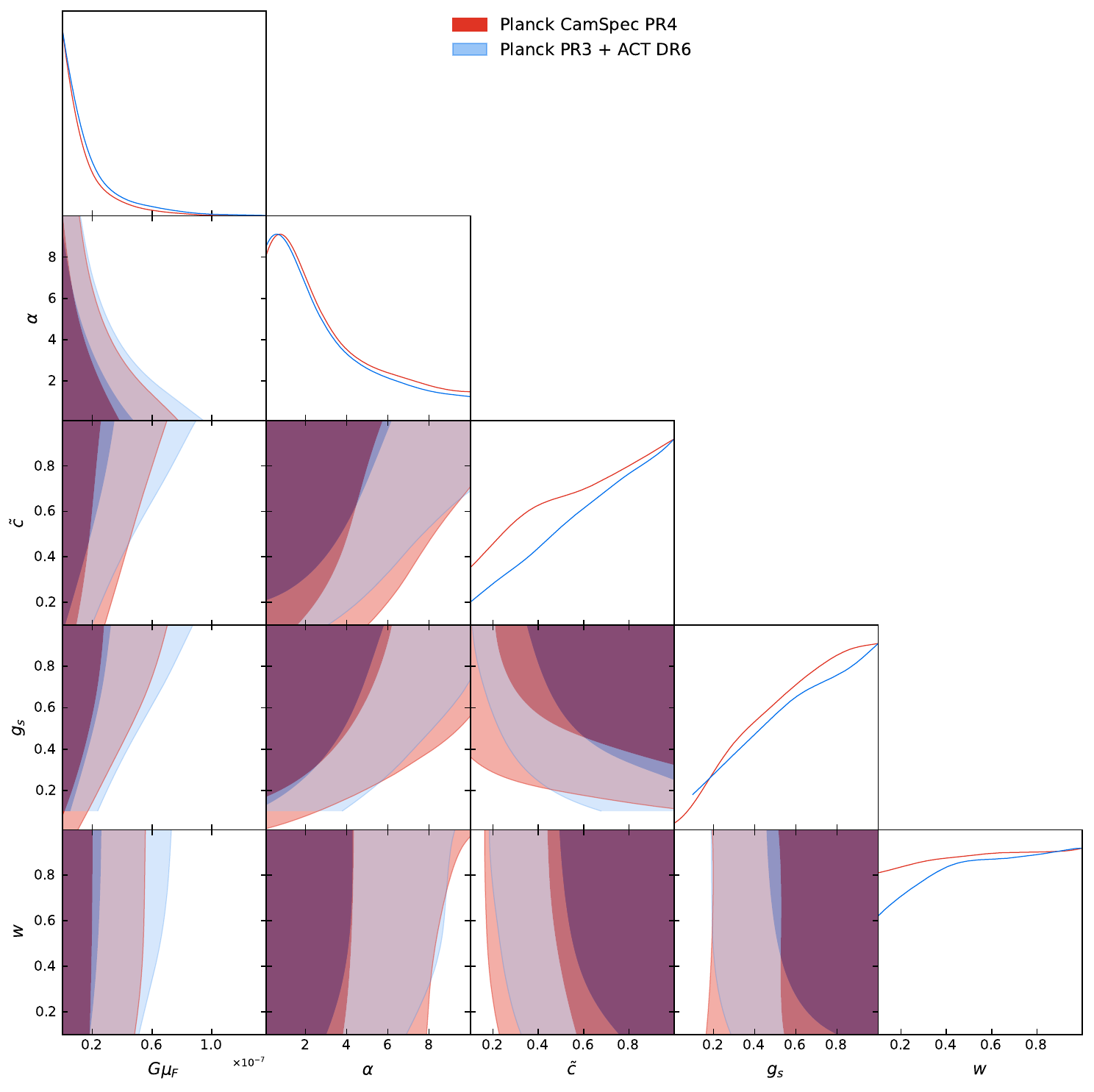}
        \caption{Linear $G\mu_F$, flat priors}
    \end{subfigure}
    \hfill
    \begin{subfigure}{0.48\textwidth}
        \includegraphics[width=\textwidth]{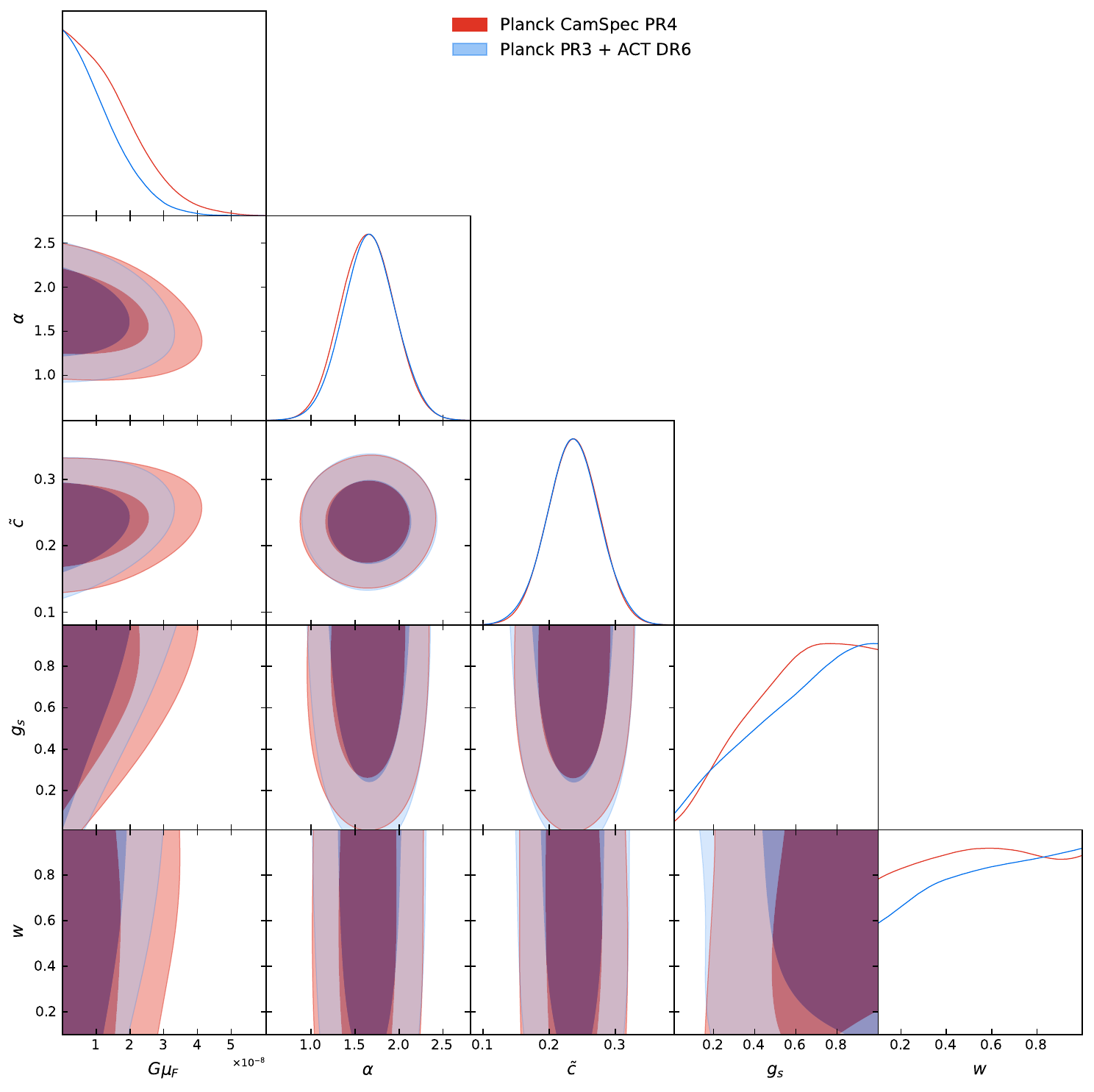}
        \caption{Linear $G\mu_F$, Gaussian priors}
    \end{subfigure}
    
    \vspace{0.5cm}
    
    \begin{subfigure}{0.48\textwidth}
        \includegraphics[width=\textwidth]{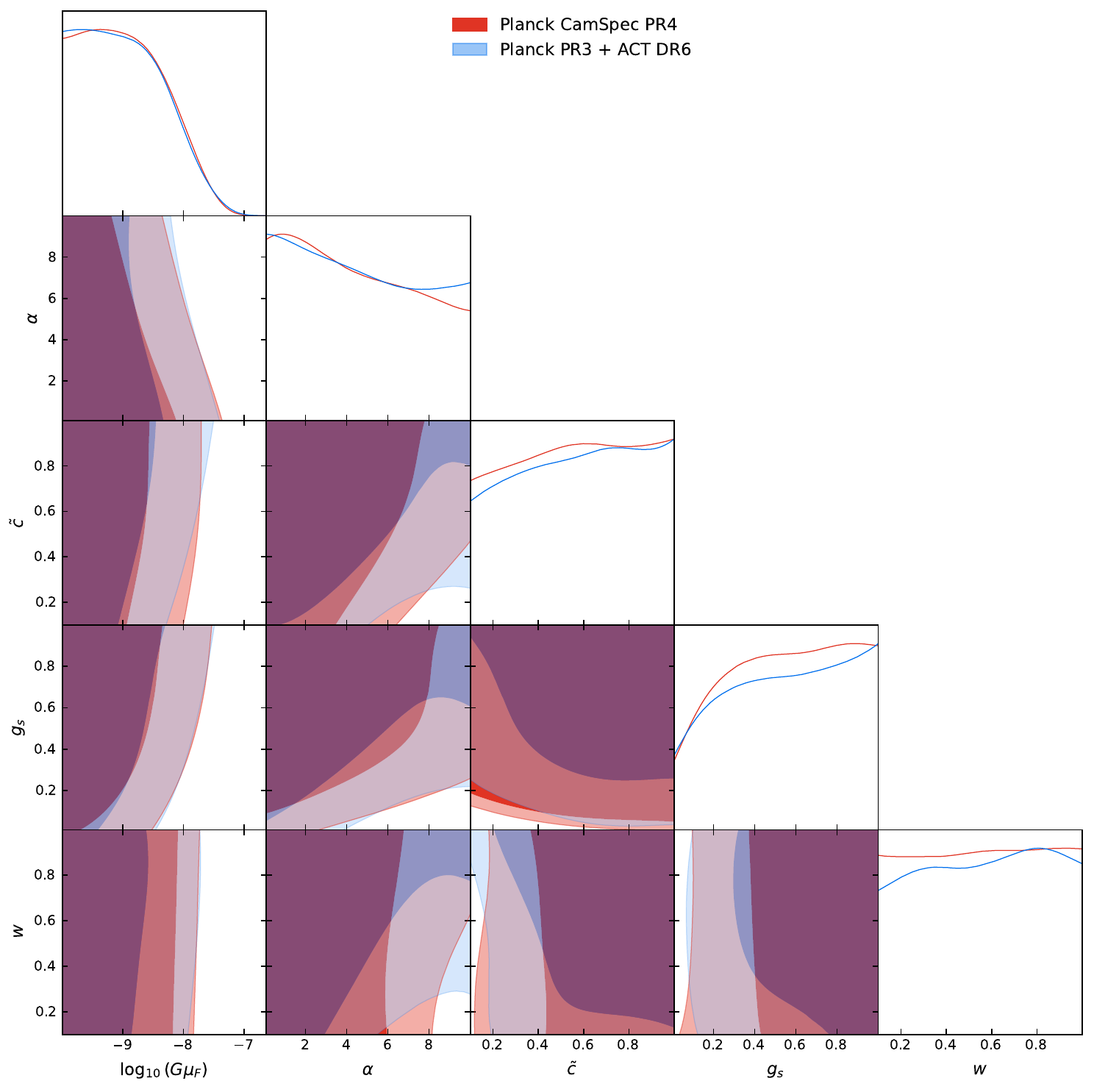}
        \caption{$\log_{10}G\mu_F$, flat priors}
    \end{subfigure}
    \hfill
    \begin{subfigure}{0.48\textwidth}
        \includegraphics[width=\textwidth]{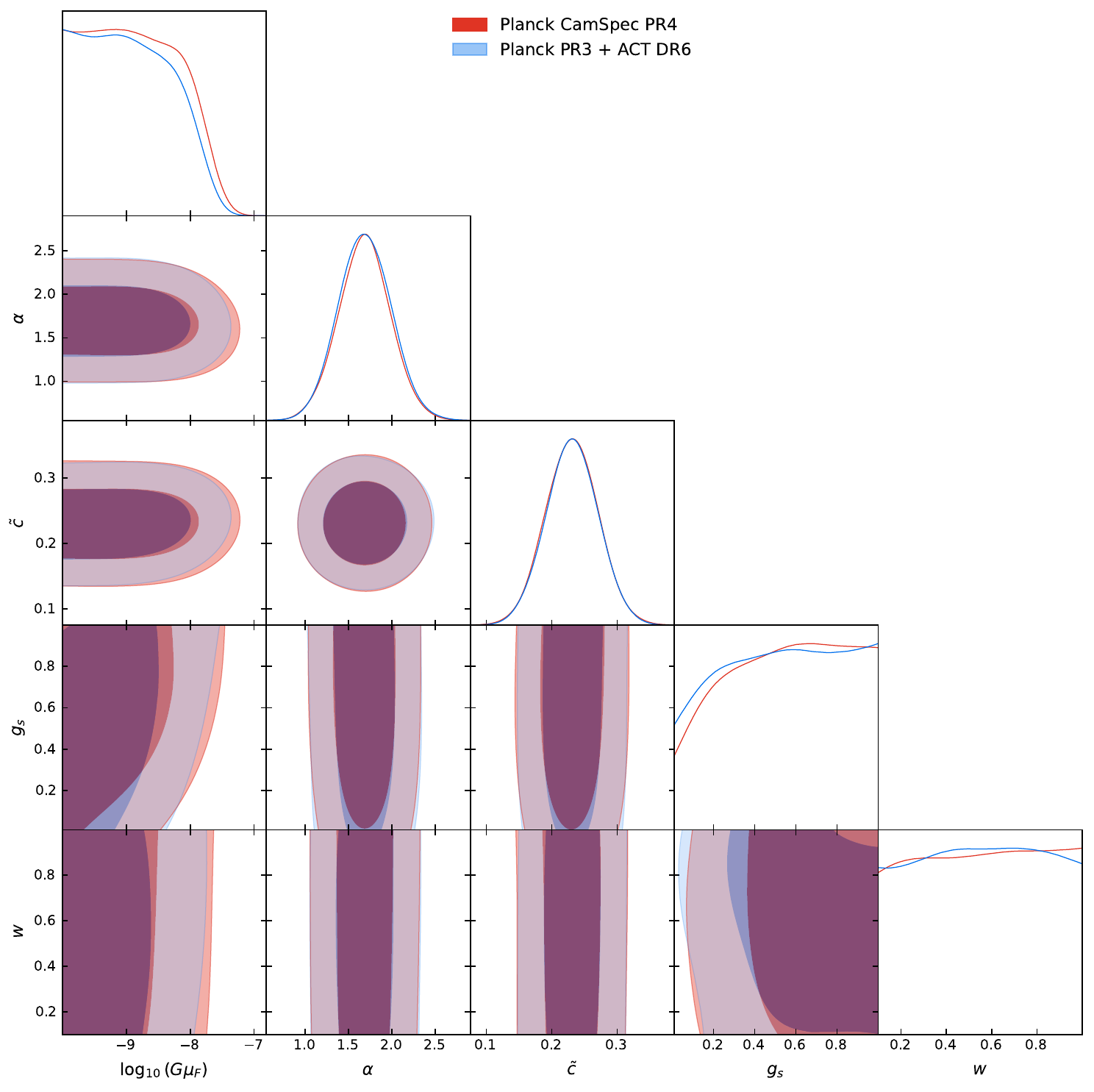}
        \caption{$\log_{10}G\mu_F$, Gaussian priors}
    \end{subfigure}
    
    \caption{Marginalized posterior distributions for the fundamental cosmic superstring parameters $\{G\mu_F, \alpha, \tilde{c}, g_s, w\}$. We compare results from \textit{Planck} CamSpec PR4 (red) and \textit{Planck} PR3 + ACT DR6 (blue). The left column (a, c) employs uninformative flat priors on $\alpha,\tilde{c}$, whereas the right column (b, d) applies physically-motivated Gaussian priors on $\alpha$ and $\tilde{c}$. While the string coupling $g_s$, effective wiggliness $\alpha$ and extra dimension volume parameter $w$ remain largely unconstrained across all cases, their inclusion modifies the shape of the tension contours.}
    \label{fig:superstringResults}
\end{figure*}

When determining Bayesian parameter constraints, the choice of priors is crucial. We perform our analysis using both a linear prior on the (super)string tension, $G\mu\in[0,2\times10^{-7}](G\mu_F\in[0,2\times10^{-6}])$, and a log-uniform prior, $\log_{10}G\mu\in[-10,-5](\log_{10}G\mu_F\in[-10,-5])$. For the string wiggliness, $\alpha$, and the loop-chopping efficiency, $\tilde{c}$, we investigate two sets of priors. Both of these parameters have accepted and regularly used simulation-derived values found in the literature, namely $\tilde{c}=0.23\pm0.04$ \cite{Martins:2000cs} and $\alpha=1.7\pm0.3$ \cite{Allen:1990tv, Pogosian:1999np, Bennett:1989yp}, where the error on $\alpha$ is chosen to cover both the matter and radiation era values derived from simulations. Our primary analysis applies the aforementioned values as Gaussian priors. We also study the scenario where an agnostic approach is taken and conservative flat priors of $\alpha\in[1,10]$ and $\tilde{c}\in[0.1,1]$ are applied to the wiggliness and chopping efficiency. For the additional superstring parameters, for which no robust constraints or preferred values exist, we apply flat priors of $w\in[0.1,1]$ and $g_s\in[0.01,1]$. Our full set of results for both types of cosmic defect across all prior and dataset combinations are summarized in Table~\ref{tab:string_constraints}. The posterior marginals for the string parameters are shown in Fig.~\ref{fig:stringResults} for cosmic strings and in Fig.~\ref{fig:superstringResults} for cosmic superstrings. The full marginals, including the $\Lambda$CDM cosmological parameters, can be found in the appendix.

\begin{table*}[t]
\centering
\renewcommand{\arraystretch}{1.5} 
\begin{tabular}{|l|l|l|c|c|}
\hline
\textbf{Model Type} & \textbf{Parametrization} & \textbf{Priors on $\alpha, \tilde{c}$} & \textbf{Planck} & \textbf{Planck + ACT} \\ \hline
\multirow{4}{*}{Cosmic Strings} & \multirow{2}{*}{$G\mu$} & Flat & $1.17 \times 10^{-7}$ & $1.16 \times 10^{-7}$ \\ \cline{3-5} 
 & & Gaussian & $6.85 \times 10^{-8}$ & $6.26 \times 10^{-8}$ \\ \cline{2-5} 
 & \multirow{2}{*}{$\log_{10} G\mu$} & Flat & $3.53 \times 10^{-8}$ & $3.53 \times 10^{-8}$ \\ \cline{3-5} 
 & & Gaussian  & $3.96 \times 10^{-8}$ & $3.66 \times 10^{-8}$ \\ \hline
\multirow{4}{*}{Cosmic Superstrings} & \multirow{2}{*}{$G\mu_F$} & Flat & $5.34 \times 10^{-8}$ & $6.55 \times 10^{-8}$ \\ \cline{3-5} 
 & & Gaussian & $3.28 \times 10^{-8}$ & $2.63 \times 10^{-8}$ \\ \cline{2-5} 
 & \multirow{2}{*}{$\log_{10}G\mu_F$} & Flat  & $1.54 \times 10^{-8}$ & $1.52 \times 10^{-8}$ \\ \cline{3-5} 
 & & Gaussian & $1.71 \times 10^{-8}$ & $1.38 \times 10^{-8}$ \\ \hline
\end{tabular}
\caption{95\% C.L. upper limits on the cosmic string tension $G\mu$ and $G\mu_F$. The ``Planck'' column uses the CamSpec PR4 likelihood, while the ``Planck + ACT'' column uses the ACT DR6 likelihood combined with a cut version of the \textit{Planck} 2018 PR3 likelihood. Results are shown across different models, parametrizations, and prior assumptions.}
\label{tab:string_constraints}
\end{table*}

Our results agree qualitatively with previous studies \cite{Charnock:2016nzm, Planck:2013mgr} -- we find no preference for either a cosmic string network or a cosmic superstring network, and all parameters aside from the string tensions remain unconstrained. This is to be expected, as the other network parameters primarily modify the overall amplitude and broad shape of the sourced spectra, rather than imprinting sharply localized features. However, the $2\sigma$ bounds on the tensions have improved significantly relative to previous work.

It is important to note the significant impact of the chosen priors and parametrizations on the final constraints. The largest differences between results arise from the linear vs logarithmic parametrization of the (super)string tension, with the $2\sigma$ tension bounds derived from a linear parametrization being consistently a factor of 2-4 larger than those derived from a log-uniform prior. This can be intuitively attributed to the prior volume effect -- a logarithmic prior assigns equal weight to every decade of tension, effectively ``pulling" the posterior toward lower values, while a linear prior gives more weight to higher tensions and yields more conservative limits.

We can quantitatively demonstrate that the discrepancy between the linear and logarithmic tension constraints is a purely statistical consequence of the chosen priors by analytically estimating the logarithmic bounds from the linear 1D marginalized likelihood. Assuming a generalized half-normal form, $\mathcal{L}(G\mu) \propto \exp[-(G\mu/s)^p]$, we fit the scale $s$ and shape $p$ parameters to the marginalized upper limits derived from our linear prior MCMC chains. By performing a change of variables, we then weight the likelihood by $1/G\mu$ to obtain an analytic prediction of the posterior that would be produced by a logarithmic MCMC simulation. In all cases where Gaussian priors were applied to $\alpha$ and $\tilde{c}$, for both cosmic strings and cosmic superstrings, the analytically estimated tension bounds are nearly identical to those obtained from the full logarithmic MCMC runs. This shows that the discrepancy can be fully attributed to prior volume effects. Note that this analytic treatment cannot be applied when flat priors are used for $\alpha$ and $\tilde{c}$, as they are degenerate with the tension and, hence, non-trivially modify the shape of the marginalized 1D tension likelihood.

We also see strong dependence on the chosen $\alpha$ and $\tilde{c}$ priors. For linear tension parametrizations, applying the Gaussian priors leads to tension bounds twice as strong as those derived using flat priors. Surprisingly, for logarithmic tension parametrizations, using Gaussian priors leads to bounds that are $\mathcal{O}(10\%)$ looser than the flat prior bounds.

A possible way to break the degeneracy between $\alpha$ and $G\mu$ is by noting that, to first order, our UETC terms scale as $(\alpha\times G\mu)^2$ suggesting we could alternatively obtain bounds on $\alpha \times G\mu$ (or $\log_{10}G\mu+\log_{10}\alpha$) with our MCMCs. By doing this, we find much closer agreement between the results obtained for the two different sets of priors on $\alpha,\tilde{c}$, with the flat prior results converging towards the Gaussian prior results. This approach does not improve the differences between linear and logarithmic tension parametrizations, as the source of this is a consequence of Bayesian statistics, as discussed above.

As prior-dependence is thought of as purely an issue of Bayesian analysis, one may hope that a frequentist approach could help solve the ambiguity in our results. To do this, we used a profile likelihood approach, where the lowest chi-squared point is found for a given tension and the resulting chi-squared profile is used to determine the $2\sigma$ bounds on the tension. While this approach is nominally prior-independent, as there is no marginalization over parameters, the choice of parameter ranges still plays a major role. The resulting chi-squared profile depends heavily on whether the ranges of the nuisance parameters are restricted to physically motivated values or allowed to vary across their full ranges. Furthermore, the technical difficulty of finding the absolute global best-fit point in a high-dimensional $\Lambda$CDM+strings parameter space can introduce sampling noise into the profile, leading to unstable bounds. In general, the profile likelihood approach favours spectra with high $\tilde{c}$ and low $\alpha$, as can be seen, for example, from the contours in Fig.~\ref{fig:stringResults} panel (a), leading to looser tension bounds than found using our Bayesian approach.

Somewhat unexpectedly, whether or not including ACT DR6 data leads to an improvement in the tension bounds is also prior-dependent. Since the $\Lambda$CDM power spectrum decays rapidly at high multipoles due to photon diffusion damping (Silk damping), whilst the spectra sourced by cosmic (super)strings fall off slower, we would expect that, at sufficiently high $\ell$, the string contribution can dominate over the $\Lambda$CDM signal and that high-resolution measurements, like those from ACT, would lead to significant tightening of tension bounds. However, as can be seen from Table~\ref{tab:string_constraints}, only bounds derived using Gaussian priors see an improvement from including ACT DR6 data, with \textit{Planck}+ACT consistently giving $\mathcal{O}(10\%)$ tighter bounds than just \textit{Planck}, regardless of tension parametrization. Note that this comparison is not strictly like-for-like, as the \textit{Planck}-only analysis uses the improved CamSpec PR4 likelihood; the improvement from ACT would be somewhat larger when compared against the same PR3 baseline.

As can be seen from the discussion above, there is significant variance between the tension bounds obtained from different prior and analysis method choices. Choosing between these approaches is largely a matter of preference, and there is no definitive way to determine which set of results represent the ``true" physical bounds. We make the choice to quote our primary results as those derived using the logarithmic parametrization with Gaussian priors on $\alpha,\tilde{c}$. These Gaussian priors incorporate our best current physical understanding of string network evolution derived from high-resolution simulations, and we have shown that, for these results, the discrepancy between bounds derived from linear vs logarithmic tension parametrizations can be fully explained as a purely statistical effect. The choice to quote the logarithmic parametrization values is made purely for a more direct comparison with the previous bounds \cite{Charnock:2016nzm}.

Finally, we note our results obtained using the \textit{Planck} 2018 likelihood are consistent with the recent \cite{Caloni:2026dyu}, where the authors combined \textit{Planck}+BICEP/\textit{Keck} data to update cosmic string tension bounds using UETC simulations with fixed $\tilde{c}=0.23$ and $\alpha=1$.

\section{Conclusions}
\label{sec:conclusions}

We have presented the first comprehensive analysis of cosmic (super)string-sourced CMB anisotropies using both the full \textit{Planck} dataset and the newly released ACT DR6 data. By combining previously derived analytic expressions for the unequal time correlators (UETCs) of cosmic (super)string networks with a novel neural-network-based emulation approach, we have constructed an efficient and accurate pipeline capable of performing full MCMC analyses of CMB anisotropies sourced by active defects. The accompanying software suite -- including the UETC calculator and the modified version of \texttt{CAMB}, \texttt{CAMBactive}, used to generate the defect power spectra -- has been made publicly available in \cite{Raidal_CAMBactive_CAMB_extension_2026}.

For both cosmic strings and cosmic superstrings, we find no statistically significant preference for models including defects over the baseline $\Lambda$CDM cosmology. All string and superstring network parameters aside from the tension remain unconstrained by the current data, including the string coupling in the superstring case. However, we obtain substantially improved bounds on the string tensions. Our primary results are obtained using physically-motivated Gaussian priors on the string wiggliness parameter, $\alpha$, and the loop-chopping efficiency, $\tilde{c}$, as these incorporate our best current understanding of string network evolution and yield results that are consistent between different tension parametrizations. For ordinary cosmic strings, we find a new $2\sigma$ upper limit of
\[
    G\mu < 3.66\times10^{-8},
\]
while for fundamental cosmic superstrings we obtain
\[
    G\mu_F < 1.38\times10^{-8}.
\]
These limits represent a significant improvement over the previous constraints \cite{Charnock:2016nzm}. Notably, we also outline the differences in results arising from the choices of parameter priors and tension parametrization, which have been ignored in previous works. The novel inclusion of the high-multipole data of ACT leads to an $\mathcal{O}(10\%)$ improvement in the bounds when Gaussian priors on $\alpha$ and $\tilde{c}$ are used.

Beyond tightening current bounds, the framework developed here demonstrates that high-resolution UETC modelling combined with fast and accurate neural emulators provides a viable and scalable strategy for analysing active sources in upcoming cosmological datasets. As future CMB experiments push toward even higher sensitivity and finer angular resolution, similar approaches will be essential for fully exploiting their constraining power on cosmic strings, superstrings, and other non-inflationary active sources of perturbations such as domain walls.

\begin{acknowledgments}

The work of AA, EJC and AM was supported by an STFC Consolidated Grant [Grant No. ST/X000672/1]. JR was supported by an STFC studentship [Grant No.\ ST/Y509437/1]. We are grateful to Jo Dunkley for encouraging us to revisit the bounds on cosmic strings and cosmic superstrings in light of the newly released ACT DR6 data. EJC is grateful to Kai Schmidt for discussions. For the purpose of open access, the authors have applied a CC BY public copyright license to any Author Accepted Manuscript version arising.

\end{acknowledgments}

\bibliography{refs}

\begin{appendix}
\section{Full posterior marginals}
\label{app:marginals}

\begin{figure*}[t]
    \centering
    \begin{subfigure}{0.48\textwidth}
        \includegraphics[width=\textwidth]{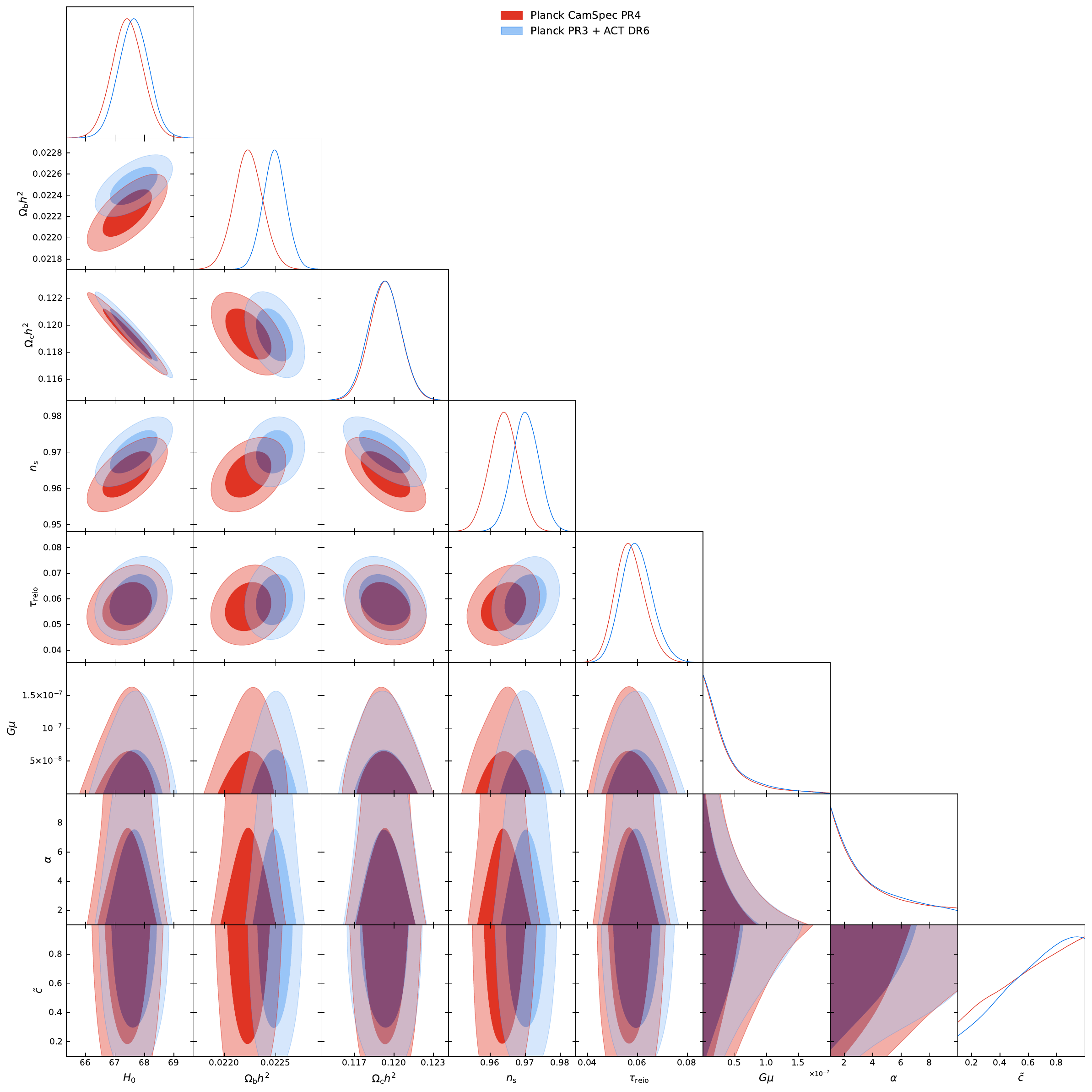}
        \caption{Linear $G\mu$, flat priors}
    \end{subfigure}
    \hfill
    \begin{subfigure}{0.48\textwidth}
        \includegraphics[width=\textwidth]{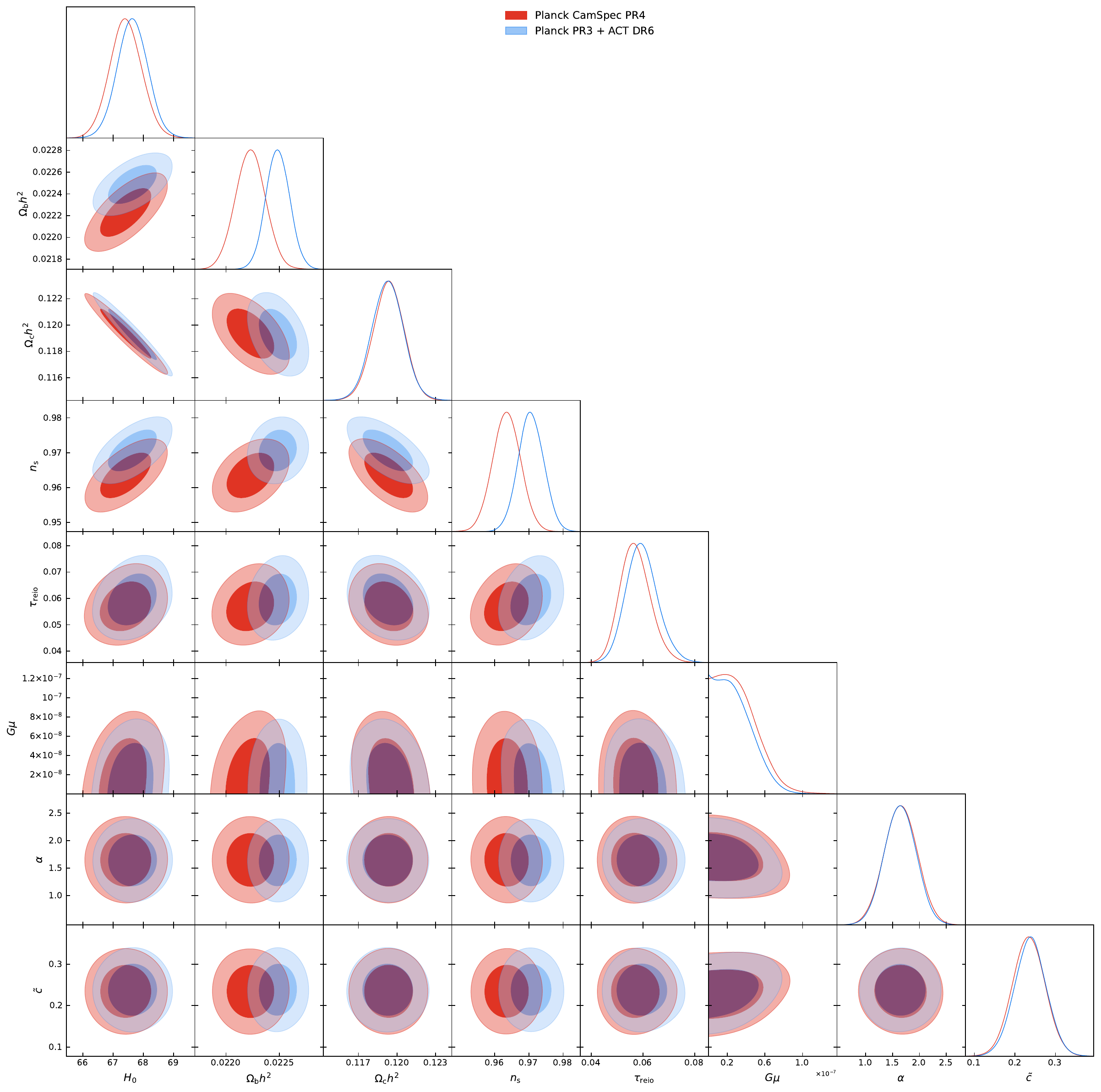}
        \caption{Linear $G\mu$, Gaussian priors}
    \end{subfigure}
    
    \vspace{0.5cm} 
    
    \begin{subfigure}{0.48\textwidth}
        \includegraphics[width=\textwidth]{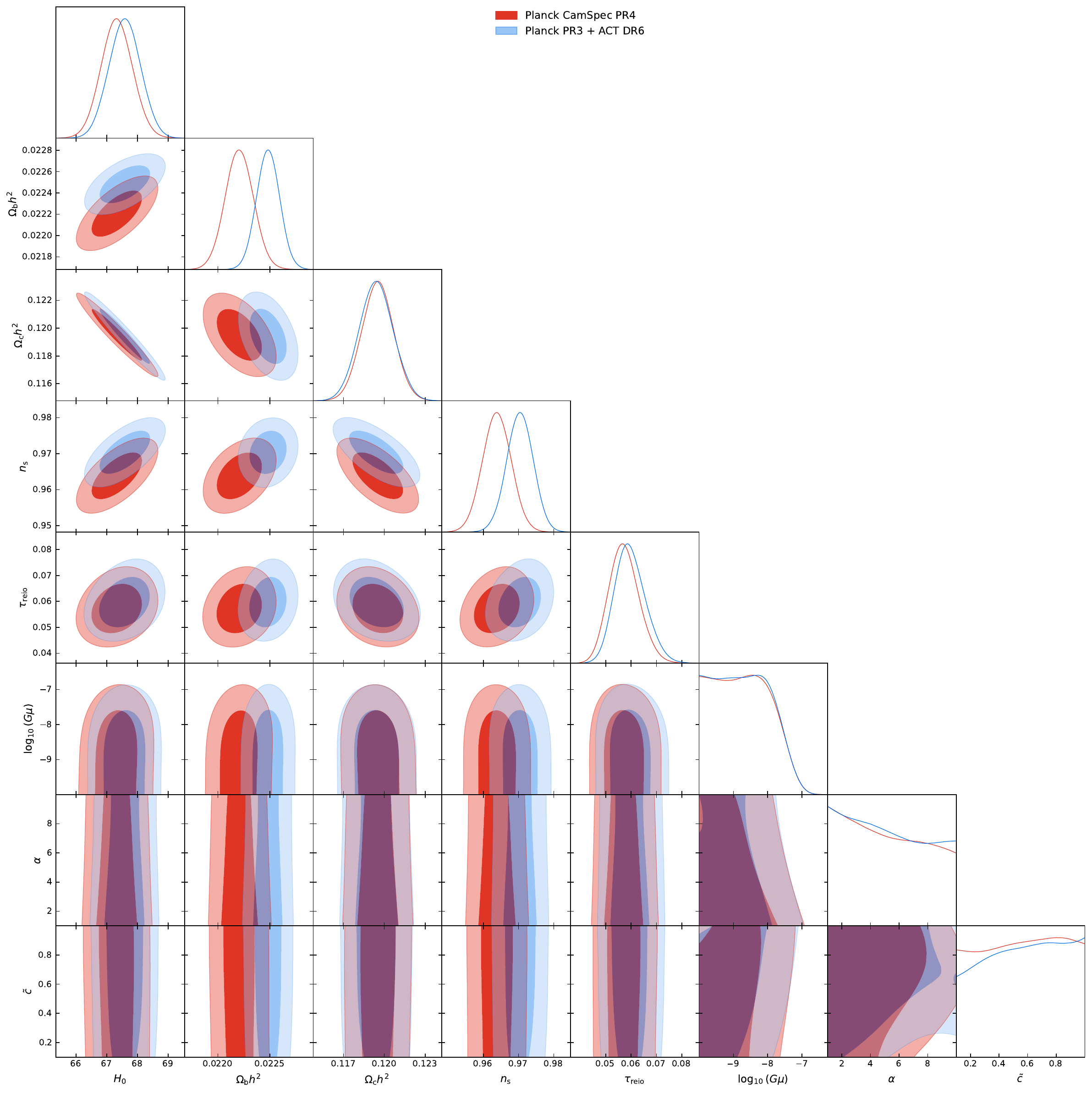}
        \caption{$\log_{10}G\mu$, flat priors}
    \end{subfigure}
    \hfill
    \begin{subfigure}{0.48\textwidth}
        \includegraphics[width=\textwidth]{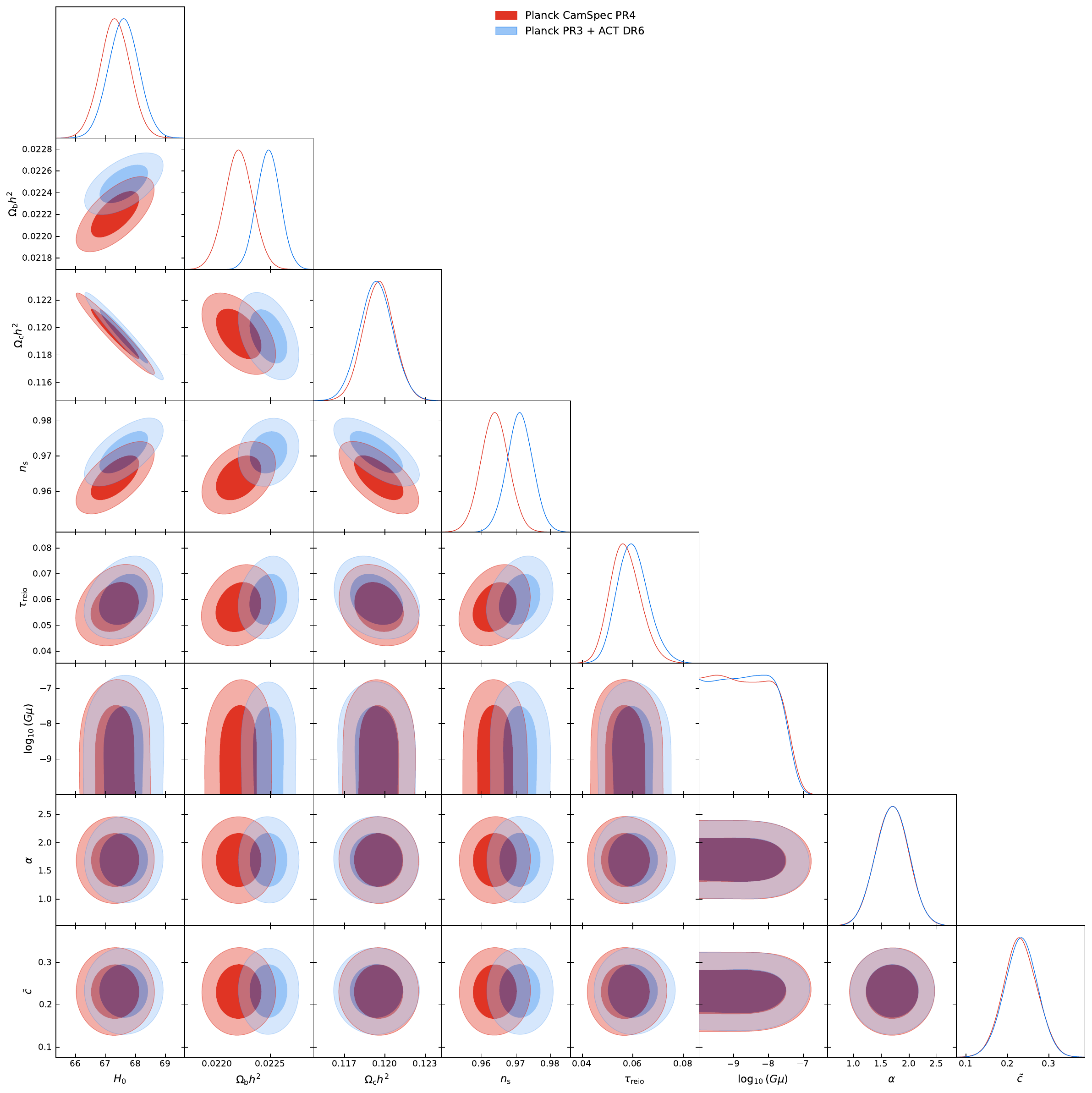}
        \caption{$\log_{10}G\mu$, Gaussian priors}
    \end{subfigure}
    
    \caption{Full set of marginalized posterior distributions for the ordinary cosmic string parameters $\{G\mu, \alpha, \tilde{c}\}$ and the cosmological parameters. In each panel, we compare the constraints from \textit{Planck} CamSpec PR4 (red) and \textit{Planck} PR3 + ACT DR6 (blue). The left column shows the results under uninformative flat priors, highlighting the prior volume effect between linear (a) and log (c) parametrizations. The right column shows the results with physically-motivated Gaussian priors on $\alpha$ and $\tilde{c}$.}
    \label{fig:stringResultsFull}
\end{figure*}

\begin{figure*}[t]
    \centering
    \begin{subfigure}{0.48\textwidth}
        \includegraphics[width=\textwidth]{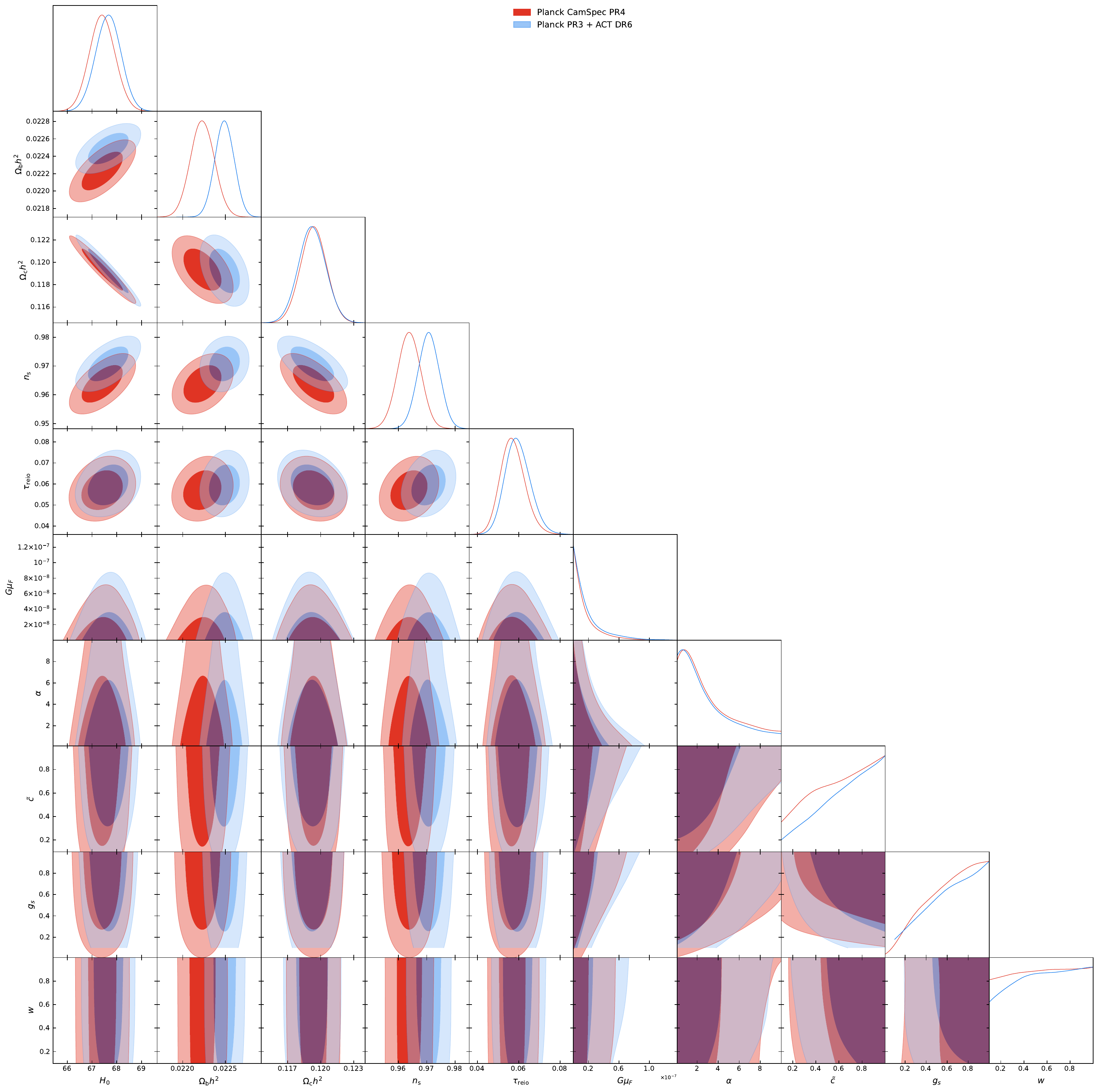}
        \caption{Linear $G\mu_F$, flat priors}
    \end{subfigure}
    \hfill
    \begin{subfigure}{0.48\textwidth}
        \includegraphics[width=\textwidth]{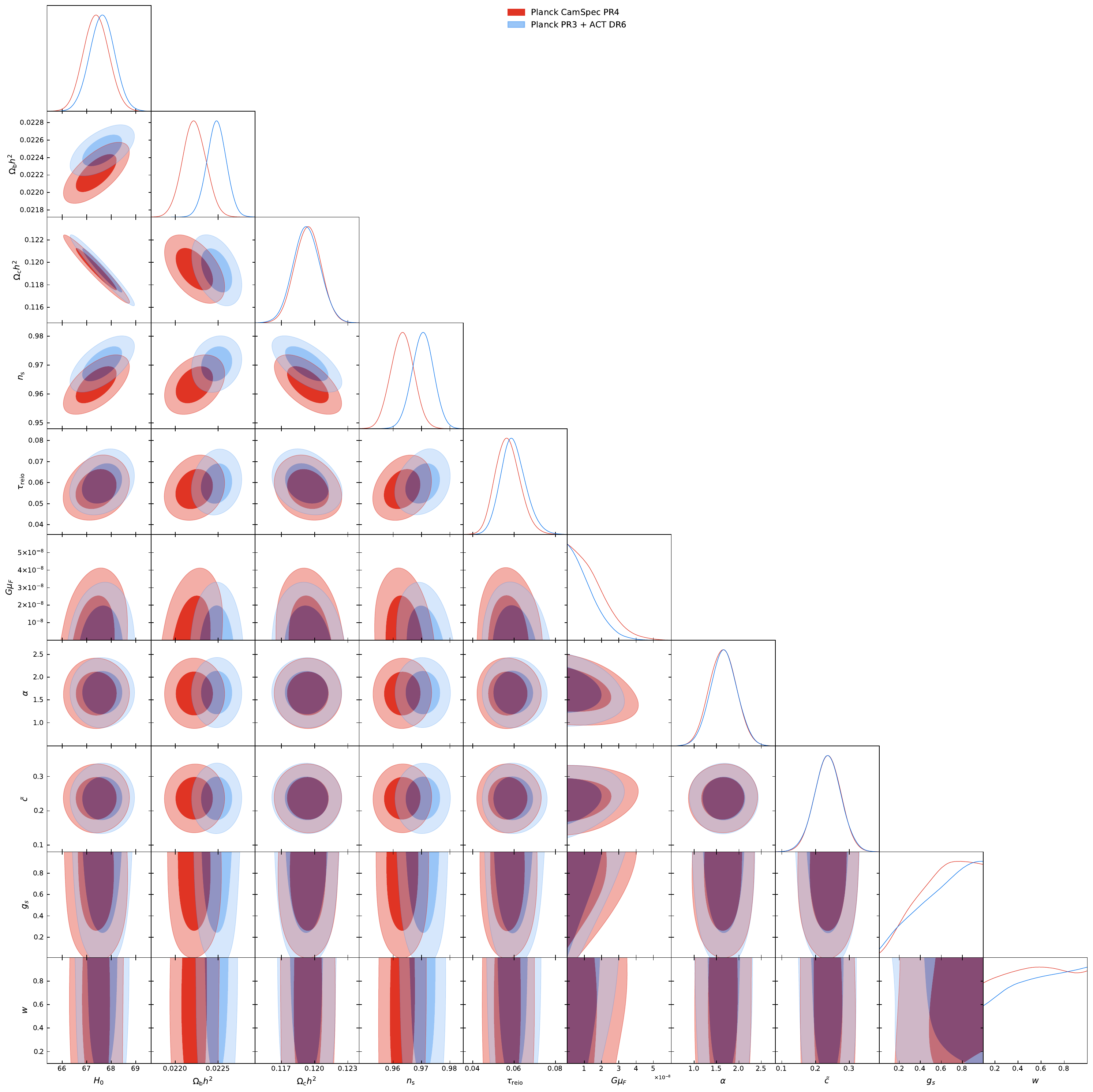}
        \caption{Linear $G\mu_F$, Gaussian priors}
    \end{subfigure}
    
    \vspace{0.5cm} 
    
    \begin{subfigure}{0.48\textwidth}
        \includegraphics[width=\textwidth]{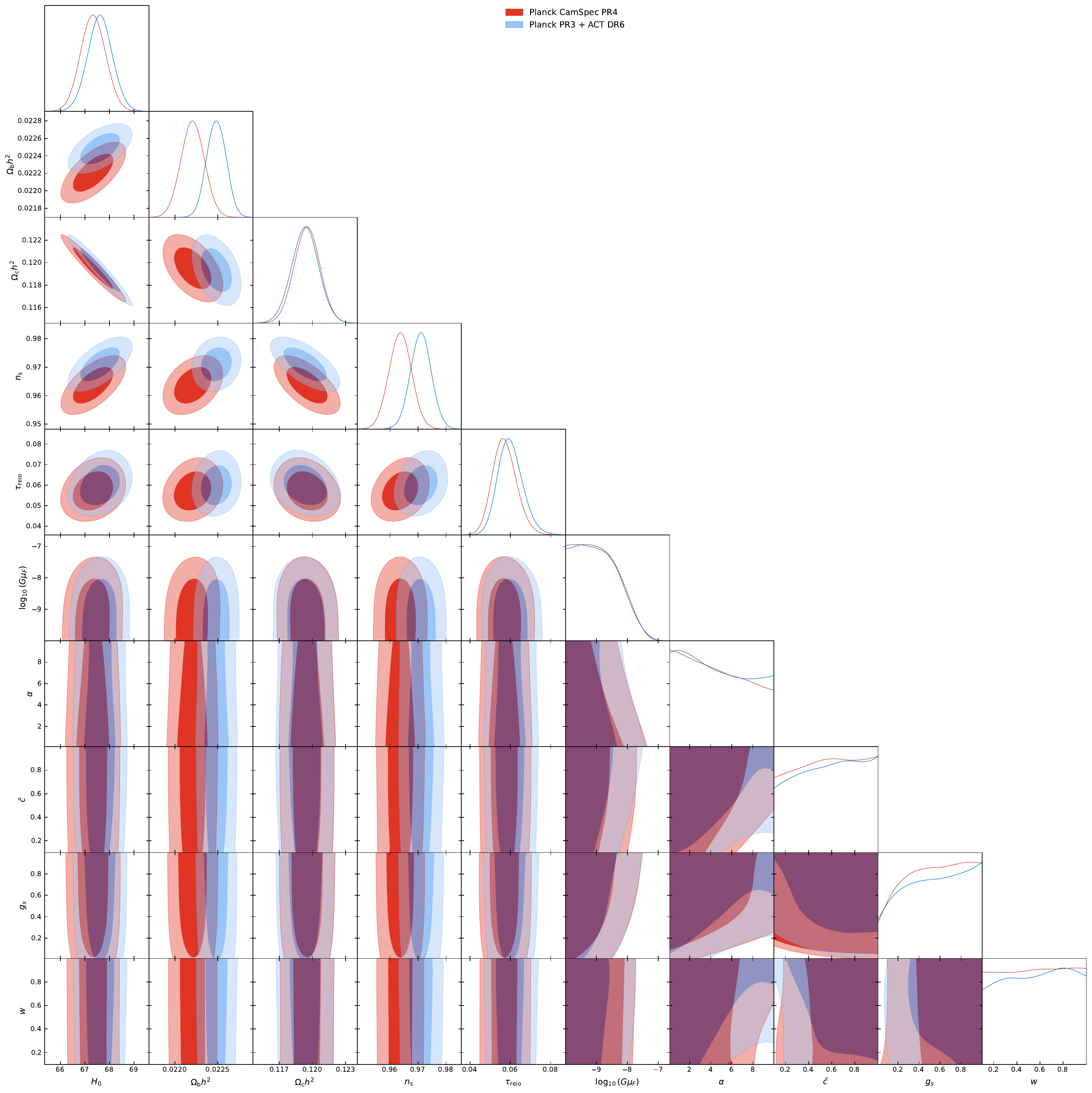}
        \caption{$\log_{10}G\mu_F$, flat priors}
    \end{subfigure}
    \hfill
    \begin{subfigure}{0.48\textwidth}
        \includegraphics[width=\textwidth]{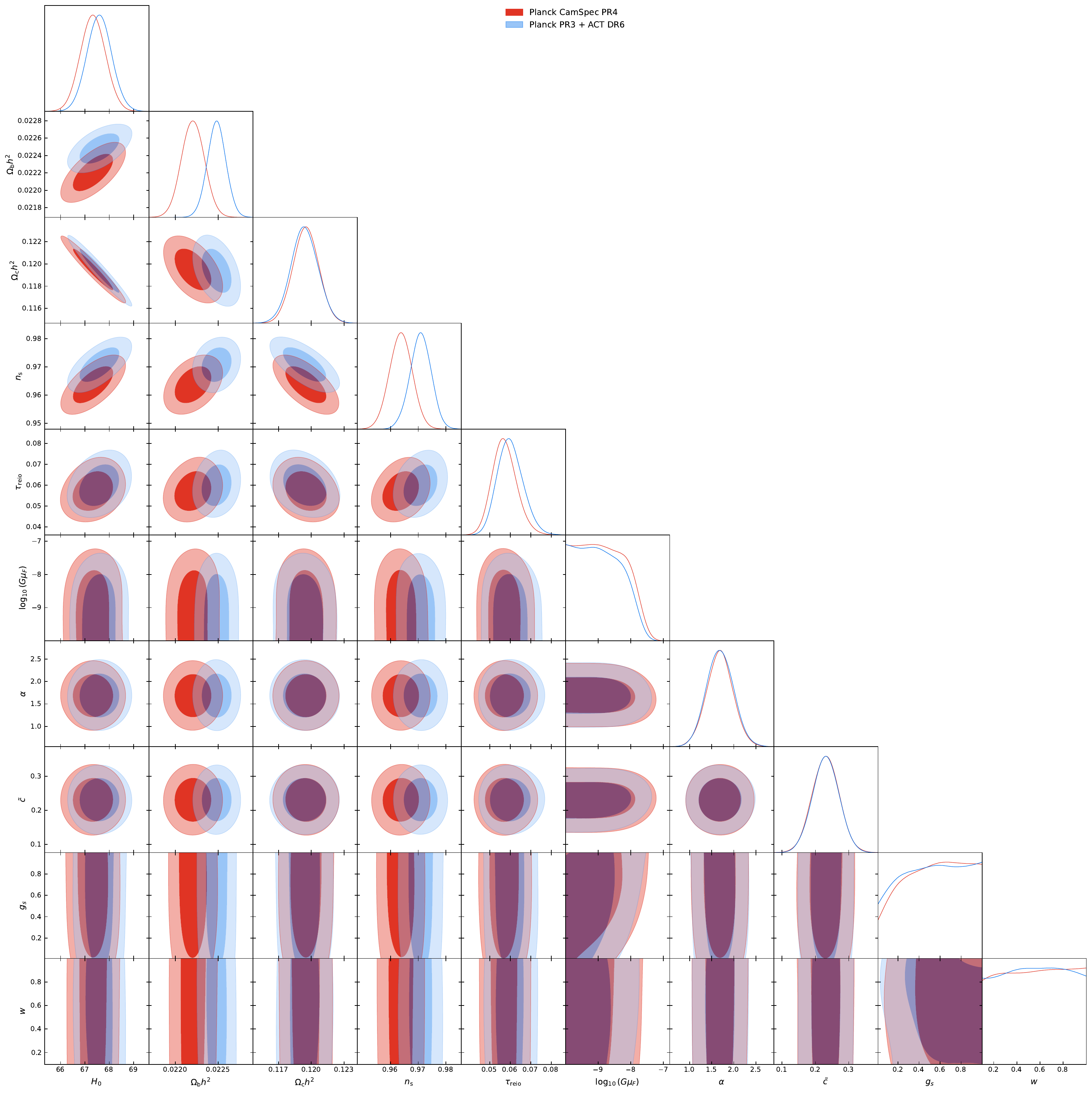}
        \caption{$\log_{10}G\mu_F$, Gaussian priors}
    \end{subfigure}
    
    \caption{Full set of marginalized posterior distributions for the fundamental cosmic superstring parameters $\{G\mu_F, \alpha, \tilde{c}, g_s, w\}$ and cosmological parameters. We compare results from \textit{Planck} CamSpec PR4 (red) and \textit{Planck} PR3 + ACT DR6 (blue). The left column (a, c) employs uninformative flat priors on $\alpha,\tilde{c}$, whereas the right column (b, d) applies physically-motivated Gaussian priors on $\alpha$ and $\tilde{c}$. While the string coupling $g_s$, effective wiggliness $\alpha$ and extra dimension volume parameter $w$ remain largely unconstrained across all cases, their inclusion modifies the shape of the tension contours.}
    \label{fig:superstringResultsFull}
\end{figure*}

\end{appendix}

\end{document}